\documentclass[twocolumn,trackchanges]{aastex63}
\usepackage{mathtools}
\usepackage{graphicx}
\usepackage{natbib}
\usepackage{amsmath}
\usepackage{booktabs}
\usepackage{listings}
\newcommand\teff{$T_\mathrm{eff}$}
\newcommand\logteff{$\log T_\mathrm{eff}$}
\newcommand\logg{$\log g$}

\graphicspath{ {./images/} }

\begin{document}

\author[0000-0001-7987-6758]{Dani Sprague}
\affil{Dept. of Computer Science, Western Washington University, 516 High St., Bellingham, WA 98225-9165, USA}

\author{Connor Culhane}
\affil{Dept. of Computer Science, Western Washington University, 516 High St., Bellingham, WA 98225-9165, USA}

\author[0000-0002-5365-1267]{Marina Kounkel}
\affil{Department of Physics and Astronomy, Vanderbilt University, VU Station 1807, Nashville, TN 37235, USA}
\email{marina.kounkel@vanderbilt.edu}

\author{Richard Olney}
\affil{Dept. of Computer Science, Western Washington University, 516 High St., Bellingham, WA 98225-9165, USA}

\author[0000-0001-6914-7797]{K. R. Covey}
\affil{Dept. of Physics and Astronomy, Western Washington University, 516 High St., Bellingham, WA 98225-9164, USA}

\author[0000-0002-5537-008X]{Brian Hutchinson}
\affil{Dept. of Computer Science, Western Washington University, 516 High St., Bellingham, WA 98225-9165, USA}
\affil{Computing and Analytics Division, Pacific Northwest National Laboratory, 902 Battelle Blvd, Richland, WA 99354-1793, USA}

\author[0000-0002-9112-9314]{Ryan Lingg}
\affil{Dept. of Computer Science, Western Washington University, 516 High St., Bellingham, WA 98225-9165, USA}
\author[0000-0002-3481-9052]{Keivan G.\ Stassun}
\affil{Department of Physics and Astronomy, Vanderbilt University, VU Station 1807, Nashville, TN 37235, USA}

\author[0000-0001-8600-4798]{Carlos G. Rom\'an-Z\'u\~niga}
\affiliation{Universidad Nacional Aut\'onoma de M\'exico, Instituto de Astronom\'ia, AP 106,  Ensenada 22800, BC, México}
\author[0000-0002-1379-4204]{Alexandre Roman-Lopes}
\affiliation{Departamento de Astronomia, Facultad de Ciencias, Universidad de La Serena.  Av. Juan Cisternas 1200, La Serena, Chile}

\author[0000-0002-1793-3689]{David Nidever}
\affil{Department of Physics, Montana State University, P.O. Box 173840, Bozeman, MT 59717-3840}

\author[0000-0002-1691-8217]{Rachael L. Beaton}
\altaffiliation{Hubble Fellow}
\affiliation{Department of Astrophysical Sciences, 4 Ivy Lane, Princeton University, Princeton, NJ 08544}
\affiliation{The Observatories of the Carnegie Institution for Science, 813 Santa Barbara St., Pasadena, CA~91101}

\author[0000-0002-5936-7718]{Jura Borissova}
\affil{Instituto de Física y Astronomía, Universidad de Valparaíso, Av. Gran Bretaña 1111, Playa Ancha, Casilla 5030, Chile}
\author[0000-0003-2300-8200]{Amelia Stutz}
\affil{Departamento de Astronom\'{i}a, Universidad de Concepci\'{o}n,Casilla 160-C, Concepci\'{o}n, Chile}
\author[0000-0003-1479-3059]{Guy S. Stringfellow}
\affiliation{Center for Astrophysics and Space Astronomy, Department of Astrophysical and Planetary Sciences, University of Colorado, Boulder,CO, 80309, USA}
\author[0000-0002-5855-401X]{Karla Pe\~na Ramírez}
\affiliation{Centro de Astronomía (CITEVA), Universidad de Antofagasta, Av. Angamos 601, Antofagasta, Chile}
\author[0000-0002-4013-2716]{Valeria Ram\'irez-Preciado}
\affiliation{Universidad Nacional Autónoma de México, Instituto de Astronom\'ia, AP 106,  Ensenada 22800, BC, México}
\author[0000-0001-9797-5661]{Jes\'us Hern\'andez}
\affiliation{Universidad Nacional Aut\'onoma de M\'exico, Instituto de Astronom\'ia, AP 106,  Ensenada 22800, BC, México}
\author[0000-0001-6072-9344]{Jinyoung Serena Kim}
\affiliation{Steward Observatory, Department of Astronomy, University of Arizona, 933 North Cherry Avenue, Tucson, AZ 85721, USA}

\author[0000-0003-1805-0316]{Richard R. Lane}
\affil{Centro de Investigación en Astronomía, Universidad Bernardo O'Higgins, Avenida Viel 1497, Santiago, Chile}

\title{APOGEE Net: An expanded spectral model of both low mass and high mass stars}

\begin{abstract}

We train a convolutional neural network, APOGEE Net, to predict \teff, \logg, and, for some stars, [Fe/H], based on the APOGEE spectra. This is the first pipeline adapted for these data that is capable of estimating these parameters in a self-consistent manner not only for low mass stars, (such as main sequence dwarfs, pre-main sequence stars, and red giants), but also high mass stars with \teff\ in excess of 50,000 K, including hot dwarfs and blue supergiants. The catalog of $\sim$650,000 stars presented in this paper allows for a detailed investigation of the star forming history of not just the Milky Way, but also of the Magellanic clouds, as different type of objects tracing different parts of these galaxies can be more cleanly selected through their distinct placement in \teff---\logg\ parameter space than in previous APOGEE catalogs produced through different pipelines.

\end{abstract}

\section{Introduction}

Effective temperature (\teff) and surface gravity (\logg) are among the most fundamental properties of a star. Determination of these parameters (or their proxy) has allowed for an understanding of how stars form and how they evolve over time.

In the early days of spectroscopy, direct measurement of these parameters was challenging due to a lack of precise stellar models. However, it was possible to assign stars a particular spectral classification that can be used as a proxy for \teff. Spectral types, ranging from O to M, were determined through examining the type and the strength of various absorption lines. Each type can further be subdivided into subtypes, usually 0 to 9. Similarly, through examining widths of certain lines, stars were assigned into different luminosity classes that can be used as a proxy of the size of a star, ranging from supergiants (I) to dwarfs (V). Hundreds of thousands of stars were classified in such a manner \citep{gray2009}. This led to complex classifications such as, e.g., G3II, to which more can be added to further express features pertaining to unique metallicity of a particular star, information on multiplicity, as well as ranges that would signify an uncertainty in classification. In the era predating computers, this allowed to immediately grasp the precise nature of a particular star of interest. Nowadays, however, given the diversity and complexity of the encoded information, such a classification can be difficult and impractical to use when dealing with a large number of stars.

Over the years, the models of stellar atmospheres have improved, allowing to create grids of synthetic spectra with known \teff, \logg, and other parameters, to which the spectra of real stars could be compared \citep[e.g.,][]{kurucz1979,kurucz1993,coelho2005,husser2013}. However, not all synthetic spectra are created equal due to inclusion of different physics, and they may have substantial systematic differences in the derived stellar parameters when matched against real data. Careful consideration of individual spectral features can help in fine-tuning the extracted parameters, but, as these features are different across different spectral types, this is not always practical to do in a self-consistent manner when analyzing spectra from large surveys.

In particular, although a good performance by various survey pipelines for programs such as APOGEE has been previously achieved for low mass stars with \teff$<$7000 K, many such pipelines have lacked calibration to measure parameters of hotter stars due to a sparsity of their spectral features \citep[e.g.,][]{garcia-perez2016}.

Previously, we have developed a neural network, APOGEE Net, which self-consistently measured \teff, \logg, and [Fe/H] for cool (\teff$<$6500 K) stars, including more exotic and rare objects, such as pre-main sequence stars \citep{olney2020}. In this work, we extend its functionality to estimate these parameters for all stellar objects, ranging from early O to late M types.

\section{Data}

\subsection{Data products}

The APOGEE project \citep[S. Majewski et al. in prep]{blanton2017,majewski2017} uses two spectrographs mounted at two 2.5 meter telescopes - one at the Apache Point Observatory (APO), and the other one at the Las Campanas Observatory (LCO) \citep{bowen1973,gunn2006}. It is capable of observing up to 300 sources simultaneously in the H band (1.51–1.7 $\mu$m), with the resolution of $R\sim$22,500 with the field of view being 3$^\circ$ in diameter at APO, and 2$^\circ$ in diameter at LCO \citep{wilson2010,wilson2019}.

The latest public data release is DR17 \citep[J. Holtzman et al. in prep]{abdurrouf2021}. It consists of $\sim$660,000 unique stars, many with multiple visit spectra. Due to the targeting strategy of the survey \citep{zasowski2013,zasowski2017}, a large fraction of the sources that have been observed are red giants. However, a number of other targets have also been observed, such as provided by additional goal and ancillary programs \citep{beaton2021,santana2021} - they include pre-main sequence stars in several star-forming regions, and main sequence and evolved stars of various spectral types. In particular A \& F dwarfs are observed in every field due to their use as telluric calibrators.

\subsection{Existing Pipelines}

\begin{figure*} 
\epsscale{1.1}
\plottwo{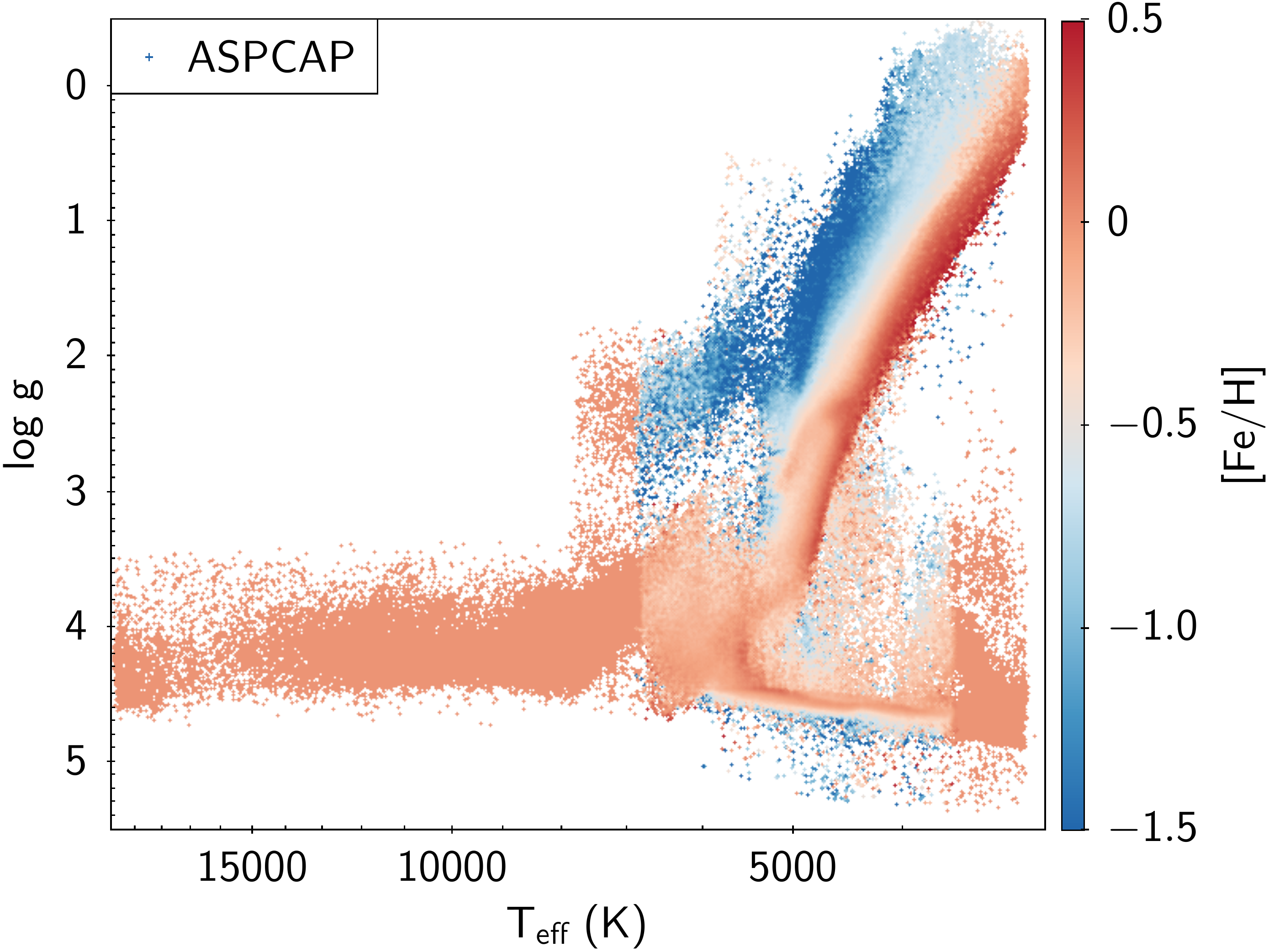}{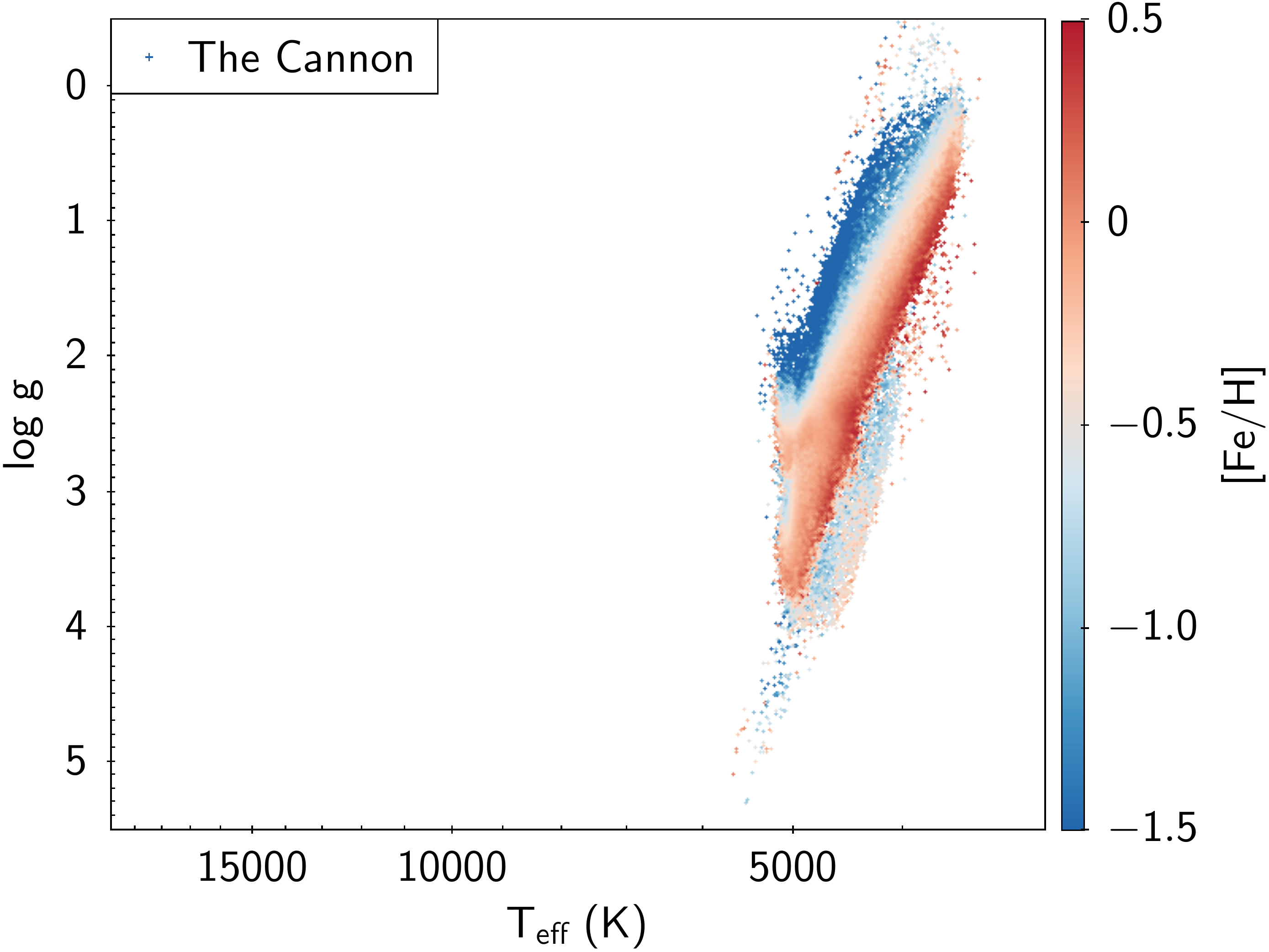}
\plottwo{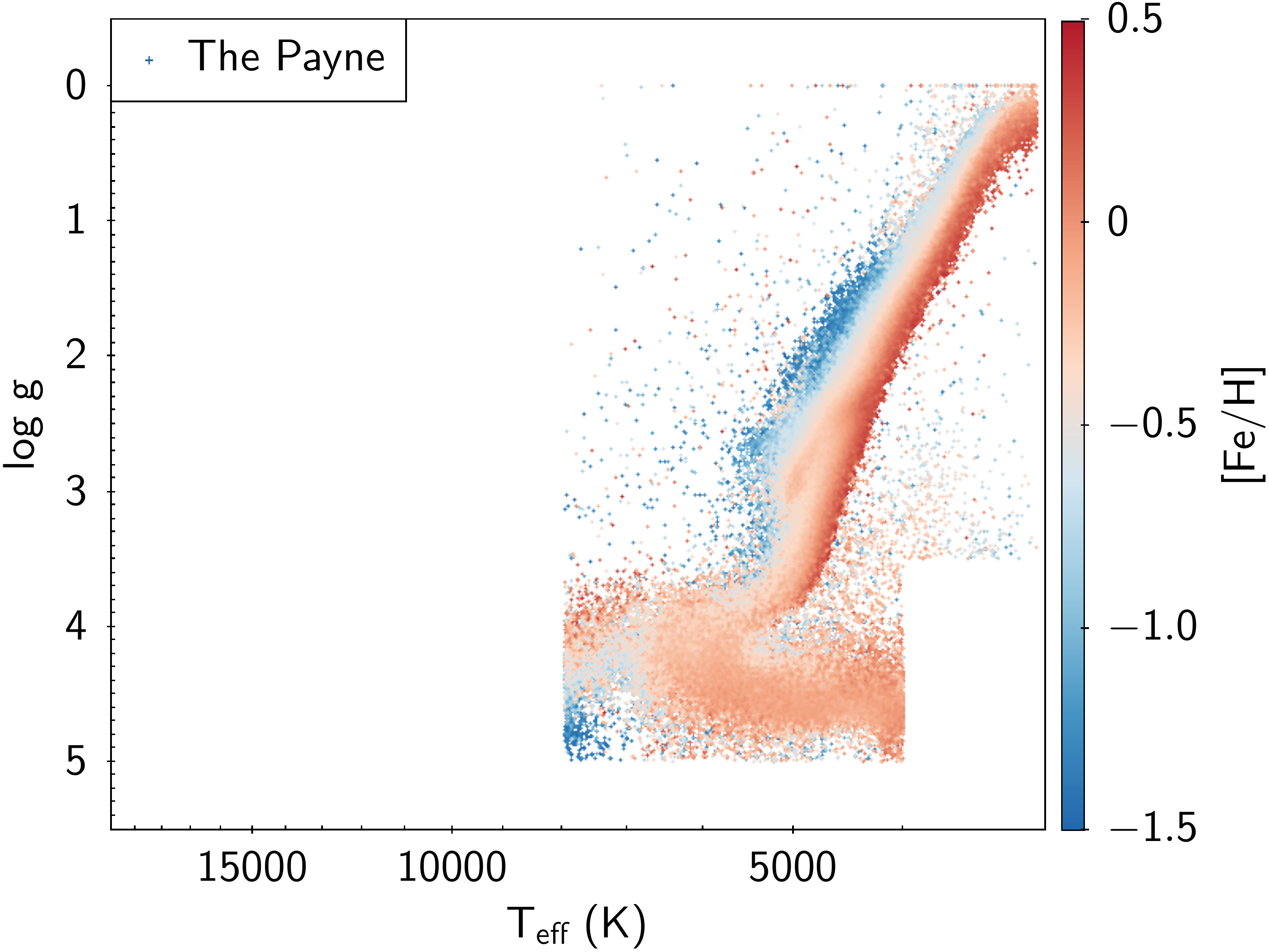}{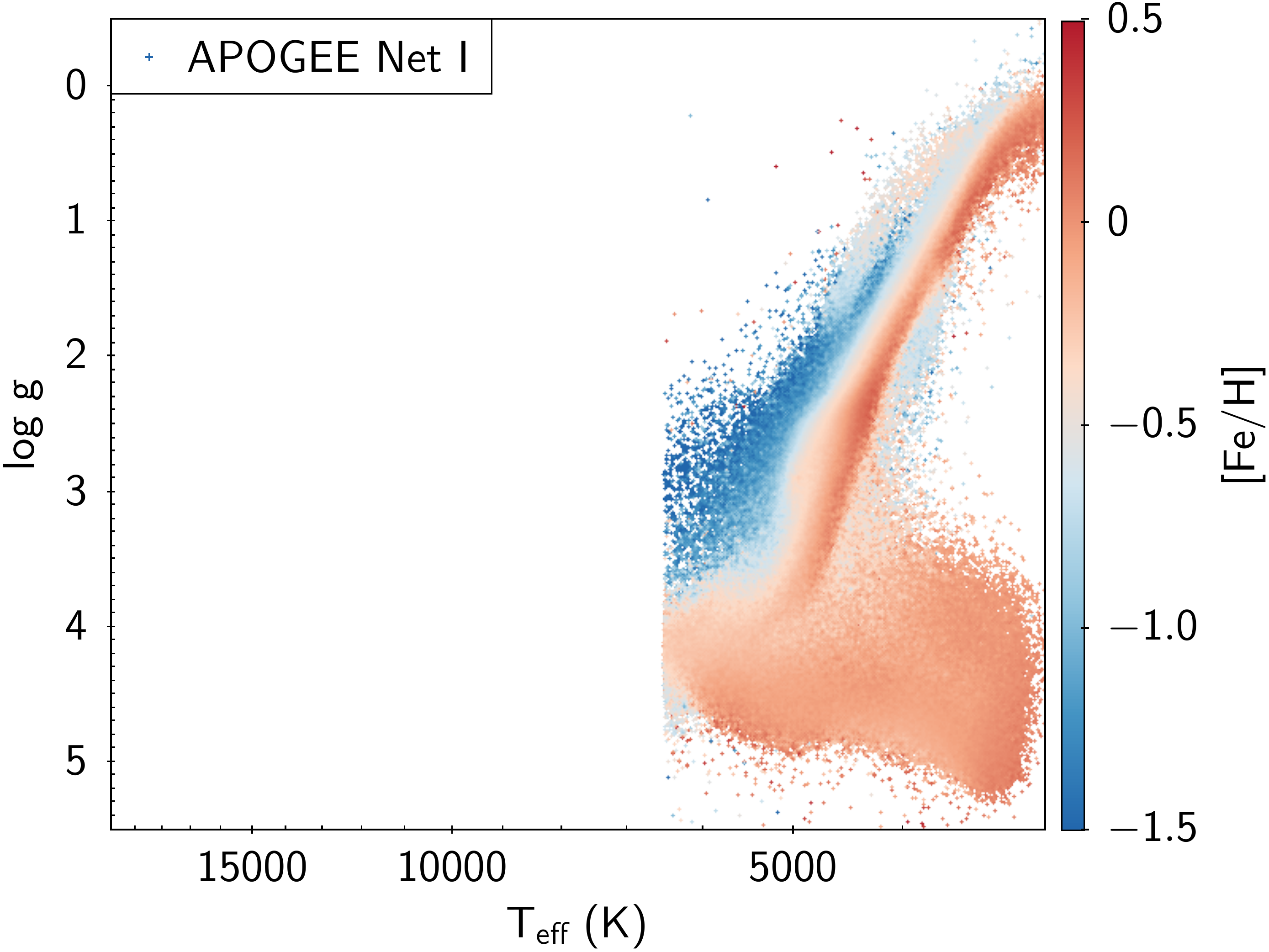}
\caption{Stellar parameters derived from APOGEE spectra from various pipeline+dataset combinations (listed clockwise from upper left): Synspec version of ASPCAP, DR17 (J. Holtzman et al. in prep; [Fe/H] is not available for cool and hot dwarfs); The Cannon \citep{ness2015}, DR14; The Payne \citep{ting2019}, DR14; and APOGEE Net I \citep{olney2020}, DR17. \label{fig:pipelines}}
\end{figure*}

The primary pipeline for the reduction and extraction of stellar parameters from the APOGEE spectra is ASPCAP \citep{garcia-perez2016}. It performs a global simultaneous fit for eight stellar parameters, which include \teff, \logg, $v\sin i$, various metallicity metrics, and other parameters, through performing a least-squares-fit between the data and a synthetic spectral grid. In old data releases, full set of derived stellar parameters were made available only for red giants. For main sequence dwarfs or pre-main sequence stars, while some \teff\ were reported, neither metallicity nor \logg\ were made available. DR16 has expanded the list of stars for which it derived full parameters to all sources with \teff$<8,000$ K \citep{ahumada2020}. DR17 (SDSS-IV collaboration in prep.; J. Holtzman et al. in prep) has added \teff\ and \logg\ for hotter stars up to \teff$<20,000$ K. However, despite the increase in coverage, and improvement in parameters over different data releases, there are some systematic features that remain which make these parameters not optimal for some types of stars, particularly those that are still young (Figure \ref{fig:pipelines}).

The first alternative to a direct grid fitting performed by ASPCAP was a data-driven approach, The Cannon \citep{ness2015,casey2016}. It was trained on a population of select stars with the parameters from ASPCAP from DR10 in an attempt to improve on some of the systematic features that arose as a mismatch between a synthetic spectra and the data. Its parameters are only limited to the red giants. The latest value added catalog was released in DR14 \citep{abolfathi2018}. From DR16 onwards it has offered no improvements compared to ASPCAP\footnote{\url{https://www.sdss.org/dr16/irspec/the-cannon/}}. There was an effort to extend The Cannon to also report \teff\ and [Fe/H] for the M dwarfs \citep{birky2020}, but there was no unified model that could perform on the full APOGEE sample.

The Payne \citep{ting2019} was another data-driven approach that was trained on the Kurucz atmospheric models. Unlike the Cannon they did not use ASPCAP parameters for label transfer, and derived their own set of labels. Prior to DR16, they were the first to produce a comprehensive set of stellar parameters (including \logg) for dwarfs with \teff$<8000$ K. Although robust across much of the parameter space, there was a degeneracy between \logg\ and metallicity among M dwarfs, confusing their parameters with pre-main sequence stars. As such, they did not report parameters for sources with \teff$<4000$ K and \logg$>3.5$ dex. 

APOGEE Net I \citep{olney2020} attempted to build on the efforts from the Payne, supplementing the available parameters for intermediate mass dwarfs and the red giants with \teff, \logg, and [Fe/H] derived from photometric relations and the theoretical isochrones for the M dwarfs and the pre-main sequence stars. This combination of the parameters was used to train a neural network that is capable of deriving stellar properties for APOGEE spectra for all stars with \teff$<$6700 K in a homogeneous manner. However, a lack of a training sample of sources hotter than \teff$>$8000 K resulted in all spectra that were dominated by the H lines to clump at the edge of the grid and, therefore, parameters for stars with \teff$>$6700 K were not reliable.

There have been numerous efforts to derive spectral types for OB stars towards a number of star forming regions through cross-matching APOGEE observations to the optical spectra with known types and deriving relations based on equivalent widths of various H lines in the near-infrared \citep{roman-lopes2018,roman-lopes2019,roman-lopes2020, borissova2019,ramirez-preciado2020}. For optimal performance, however, these efforts require initial separation of B \& F stars (as they can have a comparable strength of the H lines, requiring other lines to disentangle them). Furthermore, such a classification does not currently bridge the gap of A \& F stars to fully connect OB stars to the stellar parameters of all the other sources observed by APOGEE.

\begin{splitdeluxetable*}{cccccccBccccc}
\tablecaption{Stellar parameters for sources observed by APOGEE
\label{tab:params}}
\tabletypesize{\scriptsize}
\tablewidth{\linewidth}
\tablehead{
  \colhead{APOGEE} &
  \colhead{$\alpha$} &
  \colhead{$\delta$} &
  \colhead{\logteff\tablenotemark{$^a$}} &
  \colhead{\logg\tablenotemark{$^a$}} &
  \colhead{[Fe/H]\tablenotemark{$^a$}} &
  \colhead{SNR} &
  \colhead{\teff$_{,\mathrm{train}}$\tablenotemark{$^b$}} &
  \colhead{\logg$_{,\mathrm{train}}$\tablenotemark{$^b$}} &
  \colhead{[Fe/H]$_{,\mathrm{train}}$\tablenotemark{$^b$}} &
  \colhead{Reference\tablenotemark{$^c$}} &
  \colhead{Ref. Spec\tablenotemark{$^c$}} \\
  \colhead{ID} &
  \colhead{(J2000)} &
  \colhead{(J2000)} &
  \colhead{(dex)} &
  \colhead{(dex)} &
  \colhead{(dex)} &
  \colhead{} &
  \colhead{(dex)} &
  \colhead{(dex)} &
  \colhead{(dex)} &
  \colhead{} &
  \colhead{Type} }
\startdata
  2M00000019-1924498 & 0.000832 & -19.413851 & 3.7351 $\pm$ 0.0027 & 4.238 $\pm$ 0.038 & -0.293 $\pm$ 0.013 & 229.8 & 3.734 & 4.224 & -0.311 & Paper I                   & \\
  2M00004819-1939595 & 0.200805 & -19.666548 & 3.6667 $\pm$ 0.0024 & 1.674 $\pm$ 0.058 & -1.308 $\pm$ 0.031 & 311.7 & 3.670 & 0.88  &        & Teff, 2013AJ....146..134K & \\
  2M00011569+6314329 & 0.315407 &  63.242489 & 4.137  $\pm$ 0.020  & 3.684 $\pm$ 0.074 &                    & 238.7 & 4.306 & 4.04  &        & SpT, 1972A\&A....17..253M & B2V\\
  2M00032713+5533033 & 0.863082 &  55.550926 & 4.183  $\pm$ 0.028  & 3.81 $\pm$ 0.10 &                    & 263.2 &   &   &        &                           & \\
\enddata
\tablenotetext{}{Only a portion shown here. Full table is available in electronic form.}
\tablenotetext{a}{Parameters predicted in this work}
\tablenotetext{b}{Parameters used to train the network}
\tablenotetext{c}{Reference for the parameters used in training; SpT shows that the spectral type and luminosity class were available, Teff shows that \teff, \logg, and occasionally [Fe/H] measurements were available.}
\end{splitdeluxetable*}

\section{The APOGEE Net Model}

\subsection{Training Labels}

To construct the training labels for the high mass stars, we used SIMBAD \citep{simbad} to search for sources with existing literature measurements of their spectral parameters. Unfortunately, direct and accurate measurements of \teff\ and \logg\ for high mass stars observed by APOGEE are rare, however, many more had a reported spectral type and a luminosity class (if both are available), which can be used as a proxy of \teff\ and \logg.

To perform a transformation, we have compiled a list of high mass stars, independent of the APOGEE footprint, for which there exist independent measurement for both spectral type and luminosity class, as well as a separate measurement of \teff\ and \logg. In total, we have collated 851 measurements from \citet{lyubimkov2002,adelman2004,repolust2004,lyubimkov2005,crowther2006,kudritzki2008,martayan2008,fraser2010,lyubimkov2010,lyubimkov2012,nieva2013,bianchi2014,david2015,mahy2015,cazorla2017,martins2017,molina2018,heuser2018}.

\begin{figure*} 
\epsscale{1.1}
\plottwo{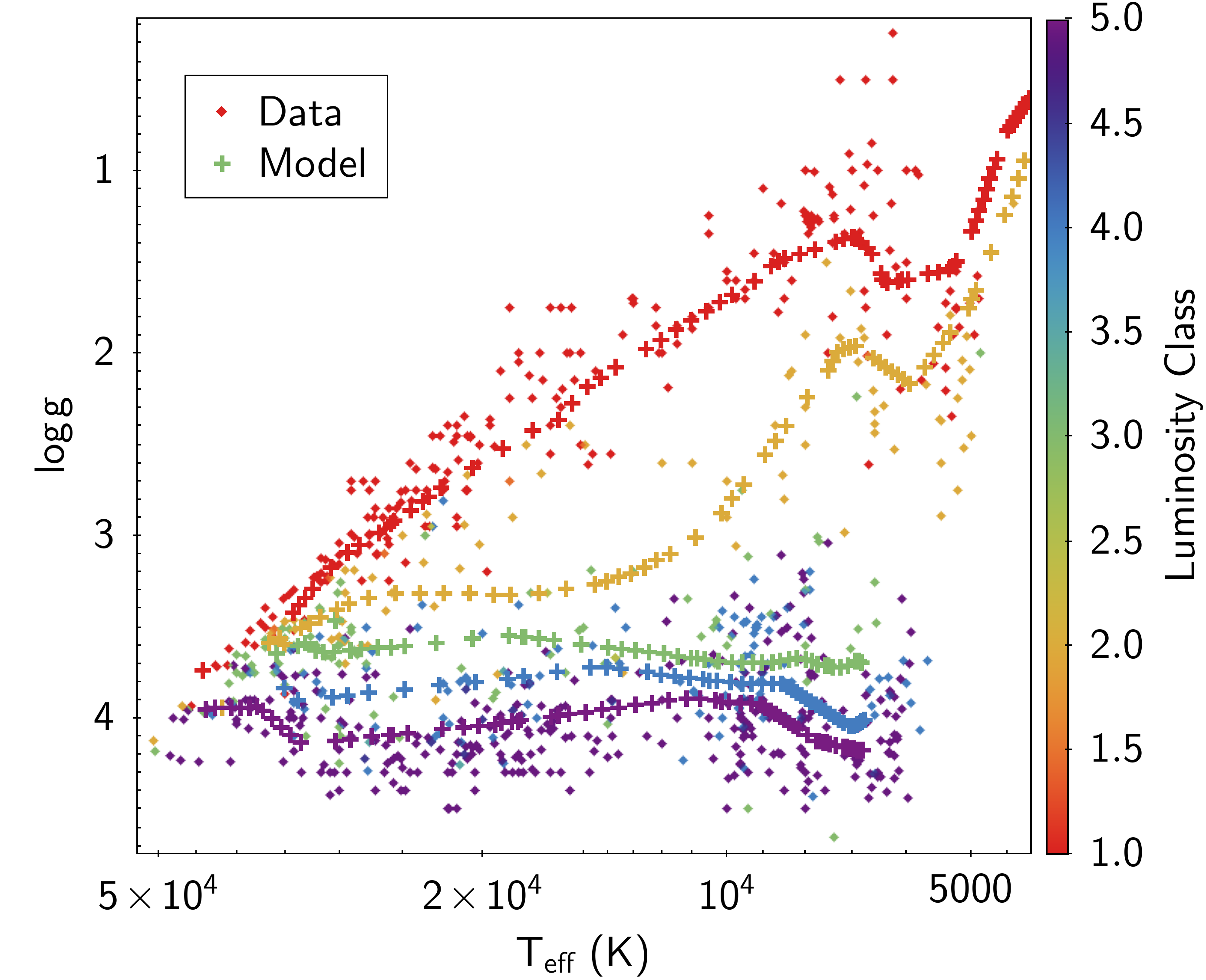}{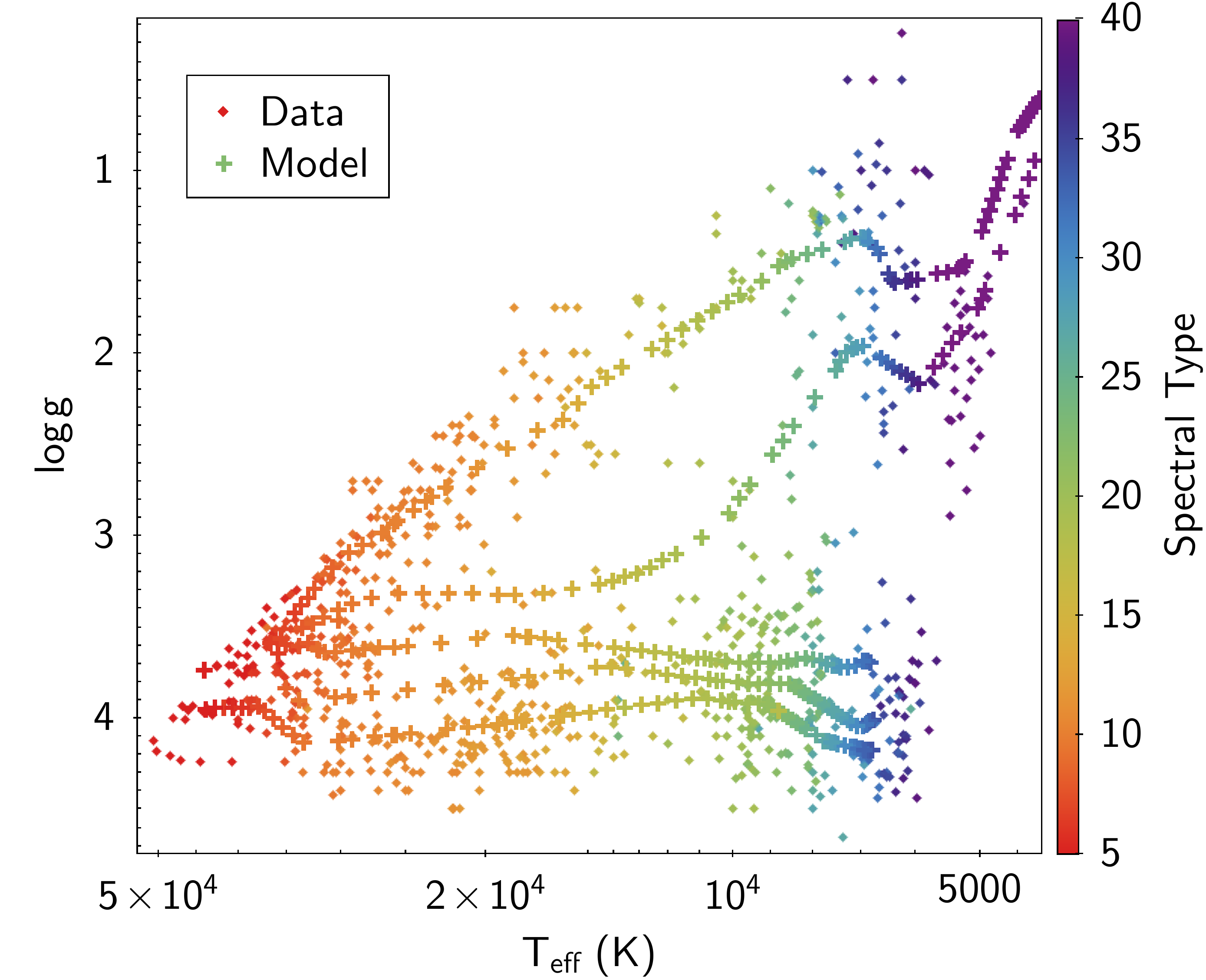}
\caption{Distribution of \teff\ and \logg\, color-coded by luminosity class (left panel) and spectral type (right panel) for sources in the literature in which both sets of these parameters are measured independently. Diamonds are the data, crosses are the computed grid. \label{fig:grid}}
\end{figure*}

We find that among OBAF stars, there is usually an unambiguous correlation between the two sets of parameters (Figure \ref{fig:grid}). We note that among the cooler stars ($<$6000 K), the luminosity class and \logg\ tend to be rather inconsistent \citep{morgan1937} -- e.g., it is relatively common for giants found in the red clump to have a luminosity class of III, IV, or V, despite having very similar \teff\ and \logg. However, as reliable \teff\ and \logg\ are available for cool stars, in their case, such an exercise of converting these \logg\ from luminosity classes is unnecessary.

We have encoded all spectral types to a numeric value: O=00, B=10, A=20, F=30, such that a star with a class of B2.5 would have a numeric label of 12.5. Similarly luminosity classes I-V were transformed to numeric labels 1-5. If a star had a range of luminosity classes listed, an average was taken -- e.g., IV-V would be labelled as 4.5. We then used a simple convolutional neural network to perform an interpolation between the two sets of parameters for the OBAF stars to construct a transformation grid.

\begin{figure} 
\epsscale{1.1}
\plotone{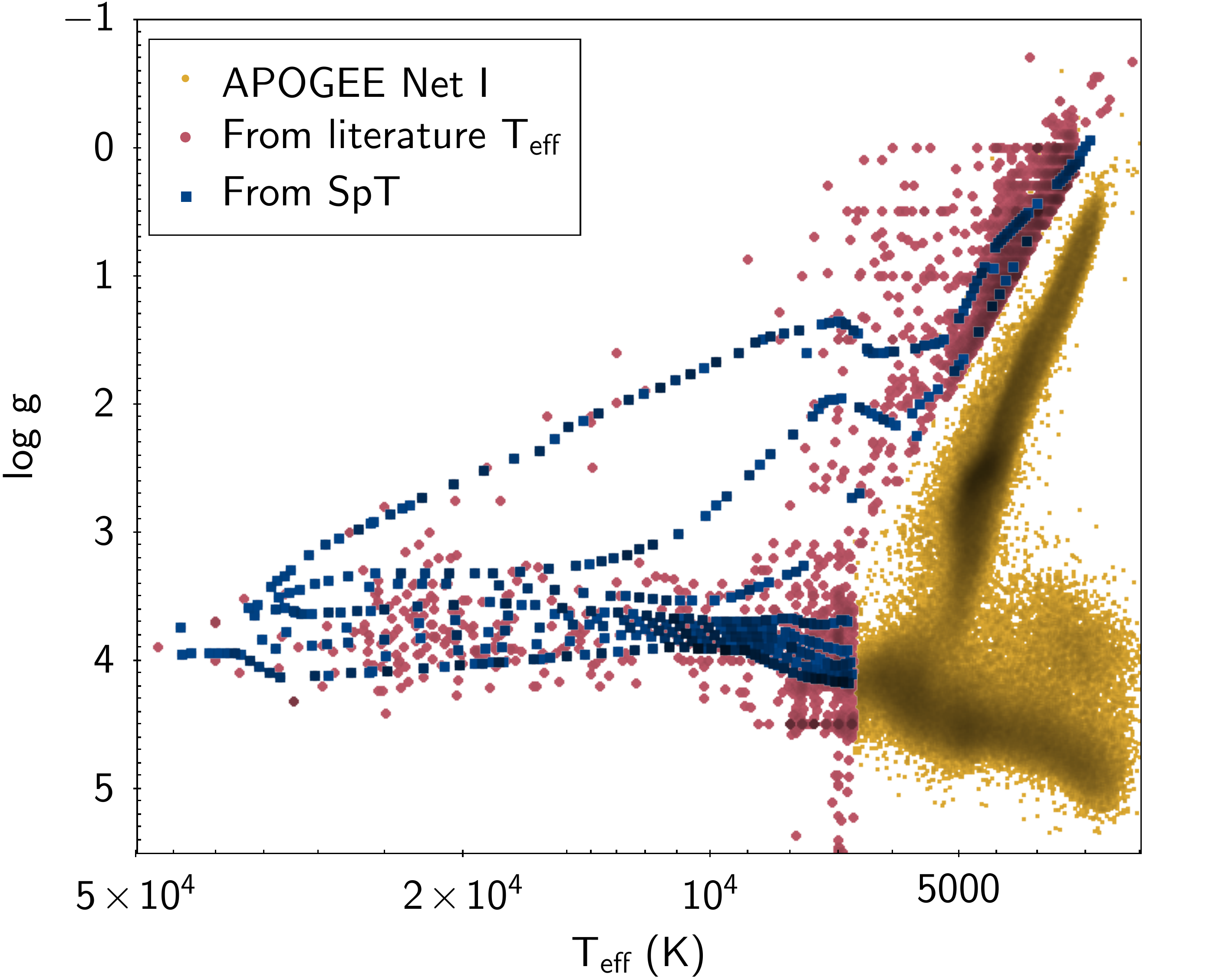}
\caption{Distribution of \teff\ and \logg\ in the training sample classes. Yellow dots are the sources from APOGEE Net I, red circles are sources with \teff\ and \logg\ from SIMBAD for hot stars in a parameter space inaccessible to APOGEE Net I. Blue squares show the sources for which \teff\ and \logg\ were transformed from the corresponding spectral type and luminosity class combination.\label{fig:train}}
\end{figure}

In total, in constructing the training sample, we retained all of the sources from APOGEE Net I if they had \teff$<$6700 K or if they had \logg$<11.3\times\log_{10}(T_\mathrm{eff})-40$ (Figure \ref{fig:train}), to potentially improve on the parameters on the supergiants, as their previously estimated \teff\ and \logg\ might have been artificially compressed due to a relative rarity of such objects. They were supplemented with \teff\ and \logg\ either transformed from the spectral type, or from independently available measurements. They are listed in Table \ref{tab:params}.

In addition to these parameters, APOGEE Net I has also included [Fe/H] as a parameter it predicts. To retain this functionality, we have preserved Fe/H input for low mass stars, and included it if an independent measurement exists. However, as a spectrum of high mass stars does not have prominent metal lines, requiring this feature would substantially limit the training sample. Thus, we have forced [Fe/H]=0 for all the hot stars where it was unavailable, and we do not report on the predicted metallicity of the stars with \teff$>$6500 K to compensate for the inclusion of these spurious values in training.

\subsection{Model Training}

\subsubsection{Initial experiment setup}

The original trained model from Paper I saw only \teff\ in the range of 3,000--6,700 K; including sources with \teff$>$50,000 K substantially skewed the weights within the network, and led to both to a decreased performance on cooler stars ($<$6000 K), as well as a lack of generalization among high mass stars.
Therefore, we instead trained a new model from scratch, using a similar model architecture and the PyTorch library \cite{pytorch}. In our new model, we converted \teff\ into $\log$ space, and renormalized \logteff, \logg, and [Fe/H] through z-score standardization. That is, given a value $x$, it is standardized as $z = \frac{x - \bar{x}}{S}$, where $\bar{x}$ denotes the mean value of our training set, and $S$ the standard deviation. The actual mean and standard deviation values are reported in Table \ref{tab:ave}.

\begin{deluxetable}{ccc}
\tablecaption{Parameters for standardization
\label{tab:ave}}
\tabletypesize{\scriptsize}
\tablewidth{\linewidth}
\tablehead{
\colhead{Parameter} & \colhead{Average} & \colhead{Standard Deviation}}
\startdata
\logteff & 3.67057 & 0.09177 \\
\logg & 2.92578 & 1.18907 \\
Fe/H & -0.13088  & 0.25669  \\
\enddata
\end{deluxetable}

\begin{deluxetable*}{cc}
\tablecaption{Classifier Hyperparameter Tuning Values
\label{tab:classparams}}
\tabletypesize{\scriptsize}
\tablewidth{\linewidth}
\tablehead{
\colhead{Hyperparameter} & \colhead{Values}}
\startdata
$Optimizer$ & SGD, ASGD, Adam, Adamw, Adamax (Best: Adamax) \\
$Learning$ $Rate$ & 0.005-0.000005 (Best: 0.00075) \\ 
$Dropout$ &  1\%-80\% (Best: 0.07646)\\
$Minibatch$ $Size$ & 128 (Best: 128)\\
$Loss$ $Weighter$ & KDE, Grid, Linear on \teff, Exponential on \teff (Best: KDE) \\
$KDE$ $Bandwidth$ & 0.08-1 (Best: 0.7671) \\
$Kernel$ $Size$ & 3, 5, 7, 11 (Best: 3) \\
$Double$ $Conv$ $Channel$ & True, False (Best: False) \\
$Disable$ $Conv$ Layers & Combination from \{ 2, 4, 6, 8, 10, 12 \} or None (Best: None) \\
$Color$ $Model$ $Depth$ & 0-10 $\in \mathbb{Z}$ (Best: 5) \\
$Color$ $Model$ $Width$ & 8, 16, 32, 64, 128, 256, 512, Varied (Best: Varied) \\
\hline\hline
\enddata
\end{deluxetable*}

The limited number of OB stars in the the training data (and even greater scarcity of blue giants and supergiants) poses challenges for training; specifically, there is a risk that the model will prioritize maximizing performance among stars in the dense regions of the \teff-\logg\ space, even if it means sacrificing performance on these relatively rare stars. To penalize the model for ignoring these rare stars, we apply a non-uniform weighting to each star in our training objective. We explored various weighting schemes for our objective function, including gridding the \teff-\logg\ parameter space and weighting stars in each grid cell inversely proportional to the number of stars in the grid cell. 
Ultimately, we settled upon a weighting scheme using Kernel Density Estimation (KDE) \cite{scott1992}. We used KDE to approximate the density of stars in the 2d standardized \teff-\logg\ space, and weight each star inversely proportional to its density.
If $d_i$ is the density estimate for a star $i$, we weight the loss on star $i$ with $\max(\frac{c}{d_i}, 5)$.  The $1/d_i$ term accomplishes the inversely proportional weighting; the max function puts a cap on how much we weight any given star; and $c$ is a constant chosen such that the average weight across stars is 1.

Training a neural network relies on particular design decisions and hyperparameters (such as the learning rate, expression for loss, architecture of the model itself, etc.). To improve the performance we tuned our model by performing a hyperparameter sweep with Weights and Biases (Table \ref{tab:classparams}), a developer tool for machine learning \citep[wandb,][]{wandb}. Tuning involves a guided search of the hyperparameter space, repeatedly training models with different hyperparameters and evaluating their resulting performance on a held out validation set.

\subsubsection{Colors}\label{sec:color}

Although the model architecture from \citet{olney2020} that operated only on the spectra could perform well on the full data sample, here we explore adding metadata for each star to further improve the performance. This metadata consists of 2MASS and Gaia EDR3 broadband photometry (G, BP, RP, J, H, K), as well as parallax \citep{2mass,gaia-collaboration2021}. 

Temperature-sensitive features can be robustly recovered directly from the spectra itself, and, as the APOGEE spectra cover the H band, the effects of reddening are not as significant in comparison to the optical light. Nonetheless, the spectra cover only a narrow $\sim0.2\mu$m band. As such, providing colors to the network allows it to infer the shape of the entire SED. Through having access to additional data, it allows the network to more finely tune its predictions.

\begin{figure} 
\epsscale{1.1}
\plotone{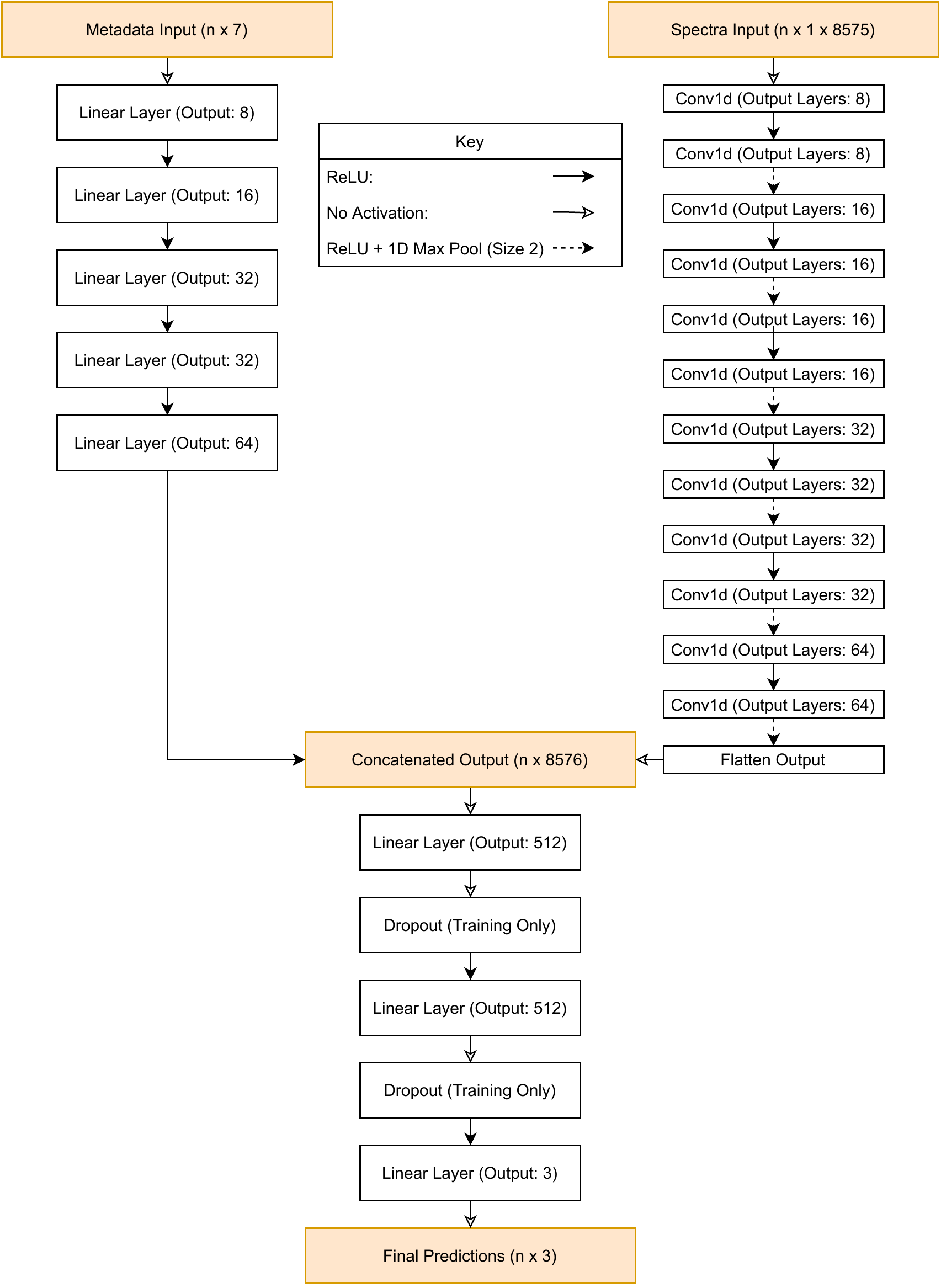}
\caption{Architecture of the neural net model used in this paper. See also Appendix \ref{sec:appendix} for code.\label{fig:model}}
\end{figure}

We feed color information as a 7-dimensional vector into a fully connected deep neural network (DNN) with rectified linear unit (ReLU) element-wise non-linearities after each hidden layer as the activation function to achieve a non-linear transformation of the data. Afterwards, the output of this DNN together with the output of convolutional layers used for the spectra is concatenated to form a single tensor. This combined tensor is subsequently passed through another DNN to generate the final output. (Figure \ref{fig:model})

When tuning the hyperparameters of this model, we also included the number of hidden layers in the DNN (from 0 to 10, where 0 meant there was no color feature branch) and the number of hidden units in each layer. 
There are alternative mechanisms one could use to inject the color feature information; we leave an investigation of their relative merits to future work. The final model is presented in Appendix \ref{sec:appendix}.

\subsubsection{Uncertainties}

We generate the uncertainties in the parameters in a manner similar as that described in \citet{olney2020}. We used the uncertainties in the spectral flux that is reported in the DR17 apStar file, and created multiple realization of the same spectrum. This was done through generating zero-mean, unit-variance random Gaussian noise for each wavelength, multiplying it by the spectral uncertainties, and adding the resulting noise to the spectrum. If the uncertainties in the spectral flux were larger than five times the median uncertainty across all wavelengths, they were clipped to that limit. This was done to prevent bad portions of a spectrum (such as in particularly noisy regions, e.g., near the telluric lines) from skewing the weights of the model. Although apStar files contain a pixel mask to identify bad pixels, it was not used, to allow the network to recognize the noise on its own and down-weight it when the model was training. The metadata (colors and parallaxes) were passed as-is without adding variance.

Different realization of the same spectrum were all independently passed through to the model, resulting in slightly different predictions of \teff, \logg, and [Fe/H]. The final reported parameters for each star are a median of 20 such predictions, which was deemed representative compared to an arbitrarily large number of scatterings for a few stars. The uncertainties in these parameters are estimated through a standard deviation, to measure a scatter in the outputs between different realizations.

\section{Results}
\subsection{Low mass stars}

The training and initial testing on a withheld sample was performed on the APOGEE data reduced with the DR16 version of the pipeline. We then apply the trained model to the APOGEE DR17 data in full. The two sets of parameters between DR16 and DR17 reductions are typically consistent with each other within the $\sim$1.2--1.5$\sigma$.

\begin{figure*} 
\epsscale{1.1}
\plotone{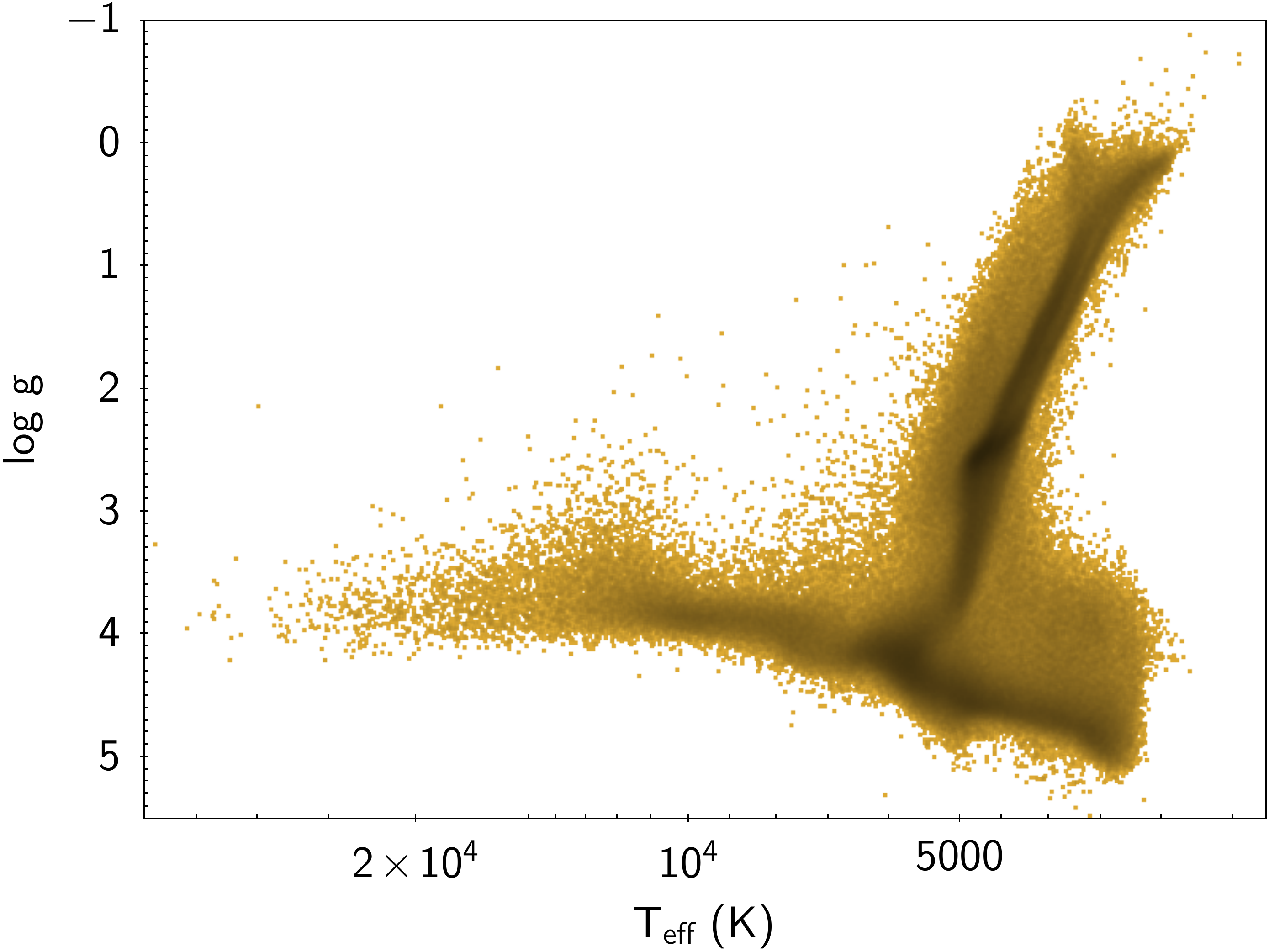}
\caption{Derived \teff\ and \logg\ distribution for all $\sim$630,000 stars in the Milky Way in the APOGEE DR17 data. \label{fig:tefflogg}}
\end{figure*}

The resulting distribution of \teff\ and \logg\ is shown in Figure \ref{fig:tefflogg}. Although [Fe/H] are predicted for all stars, we do not report it for sources with \teff$>$6500 K, as the training labels for them are unreliable.

The typical reported uncertainties are 0.03 dex in \logg, 0.002 dex in \logteff\ (30 K for a 6000 K star), and 0.01 dex in [Fe/H], which is largely consistent with the typical uncertainties in APOGEE Net I. On average the reported uncertainties are also comparable to those reported by ASPCAP DR17 for these parameters.

\begin{figure*} 
\epsscale{1.1}
		\gridline{\fig{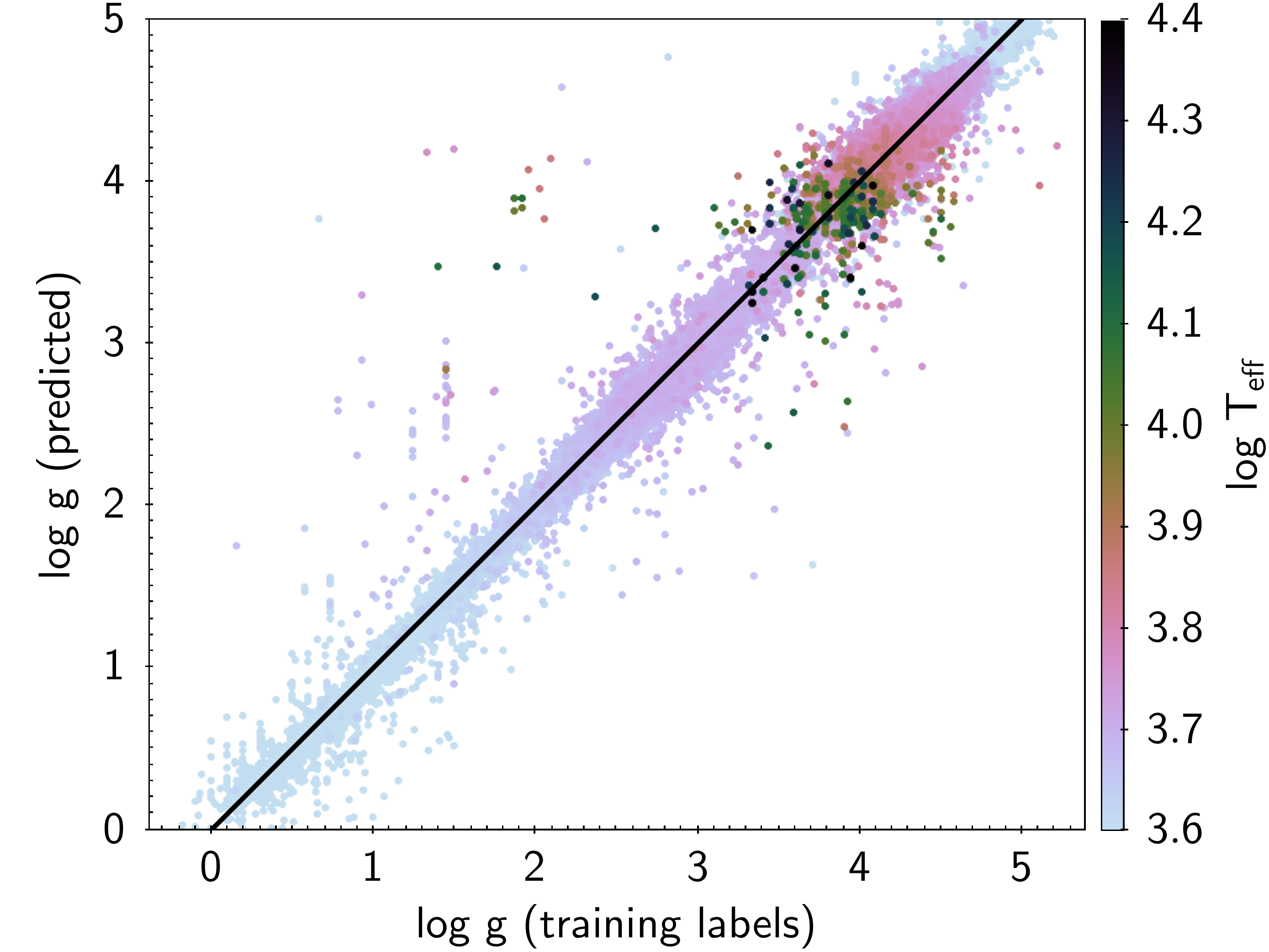}{0.33\textwidth}{}
             \fig{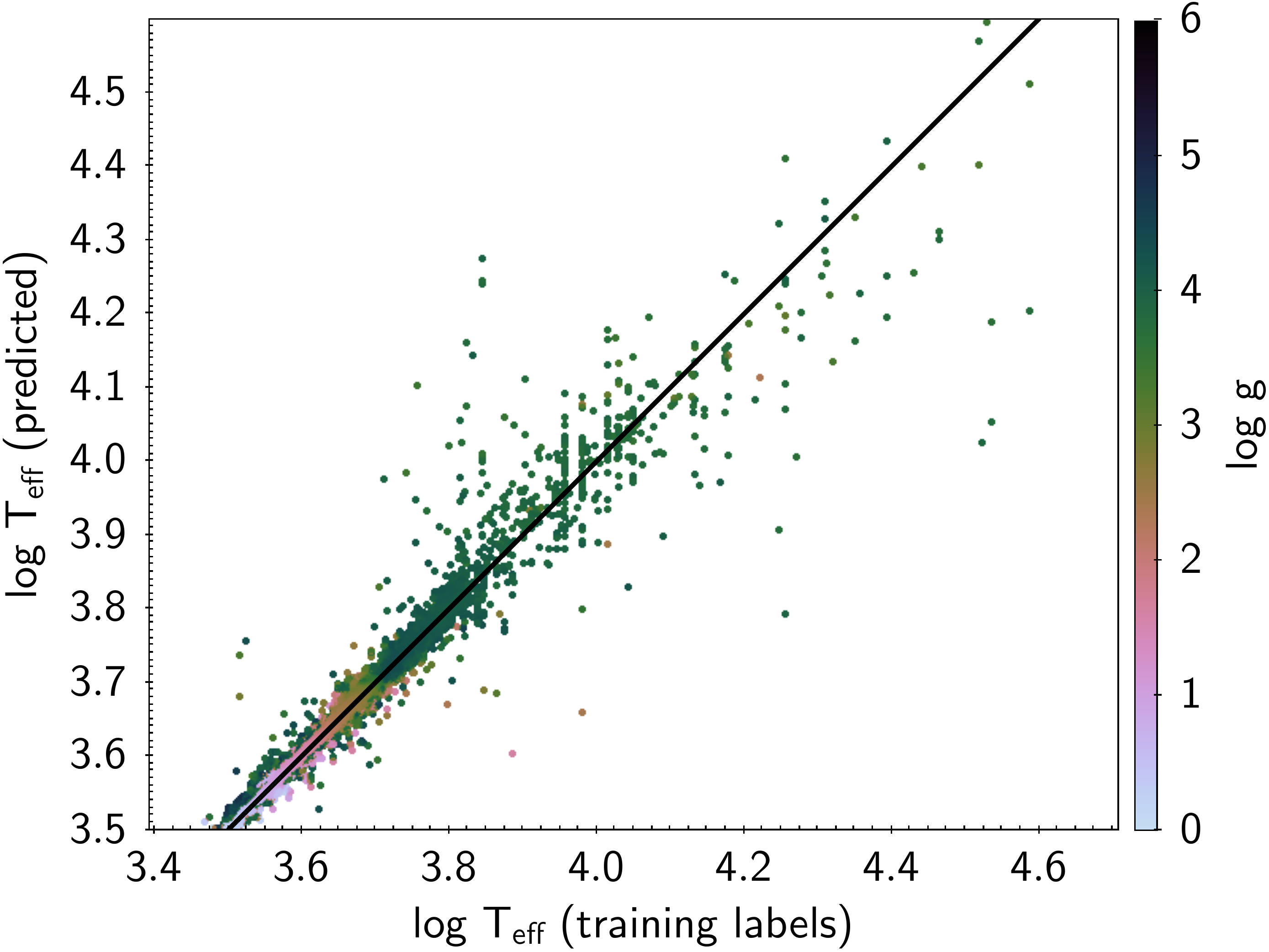}{0.33\textwidth}{}
             \fig{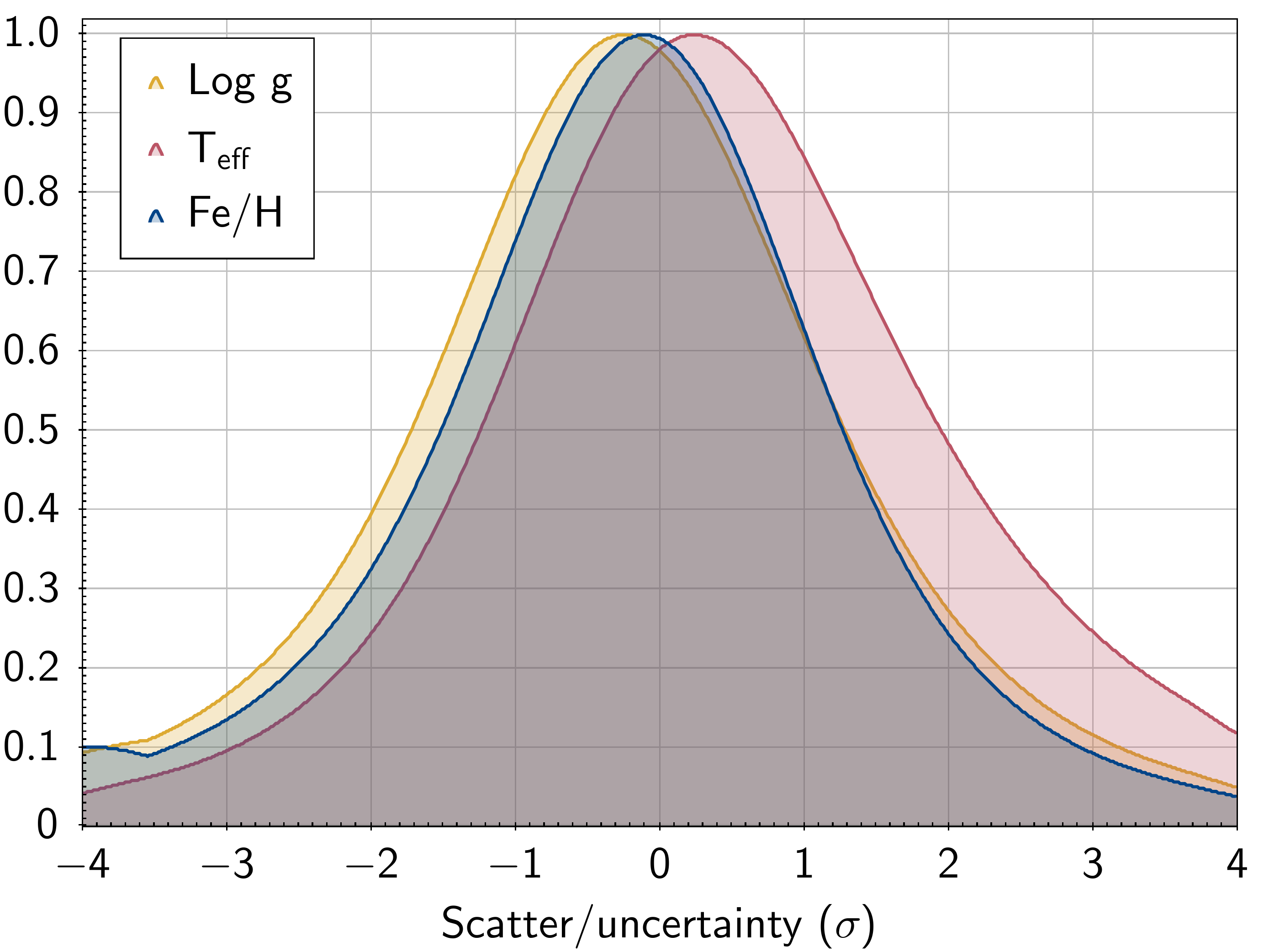}{0.33\textwidth}{}
        } \vspace{-0.8cm}
\caption{Comparison of parameters between the training labels and the model predictions, using the reserved test sample that was not used in training the model. Left: Comparison of \logg, color coded by \teff. Middle: Comparison of \teff, color coded by \logg. Right: Difference between the training labels derived from APOGEE Net I for cool stars with \teff$<6500$K and the model predictions for the same stars in this work, divided by the uncertainties measured both by APOGEE Net I and II added in quadrature, to demonstrate the typical magnitude of residuals, in $\sigma$. \label{fig:onetoone}}
\end{figure*}

Overall, the parameters for cool stars show a good agreement with APOGEE Net I (Figure \ref{fig:onetoone}). Examining the difference in the parameters between two versions of the pipeline relative to the reported uncertainties, the scatter is typically within 1.5-2$\sigma$ (Figure \ref{fig:onetoone}, right panel). As such, this is likely a factor that should be considered regarding a systematic uncertainty in the absolute calibration of the spectral parameters between two separate models.

While the derived parameters with APOGEE Net I and II are largely self-consistent, there may be further systematic features that may not necessarily be accounted by the model. Different parts of the parameter space had different means of deriving their initial set of labels. Some of those labels were based on synthetic spectra matching to the data, some of them were based on theoretical isochrones, and some of them were based on empirical photometric relations. Such a stitching together of various sets of labels may have caused discontinuities in the parameter space that the model has learned to interpolate across. The physics included in computing the original set of synthetic spectra from which some of the labels were derived may be incomplete - this may introduce systematic offsets in the predicted parameters that are difficult to quantify. This is generally the case with most pipelines deriving parameters for data from large spectroscopic surveys. Nonetheless, the accuracy that can be inferred from the overall parameter distribution is sufficient for a wide variety of applications.

\subsection{High mass stars}

\begin{figure*} 
\epsscale{1.1}
\plottwo{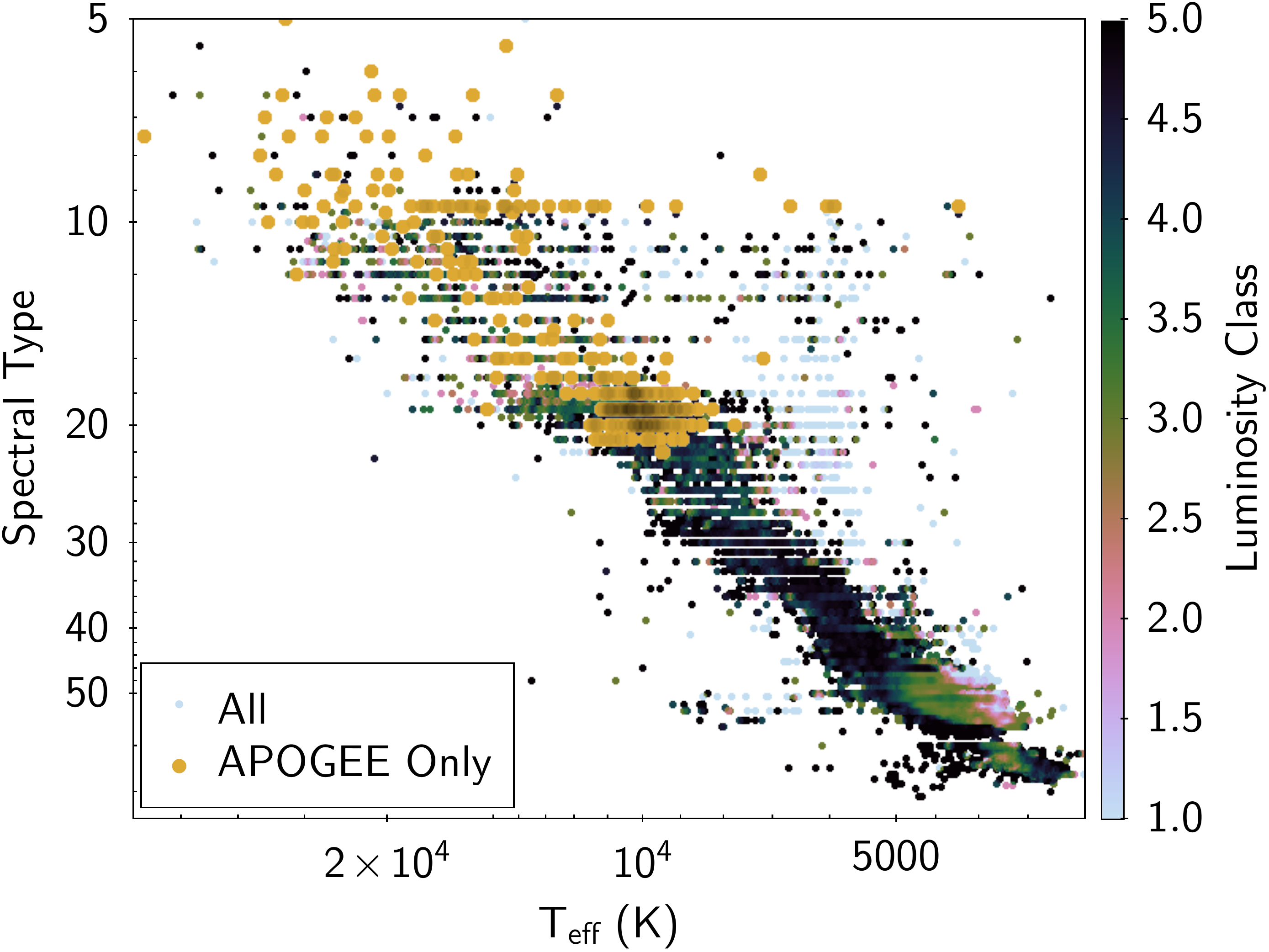}{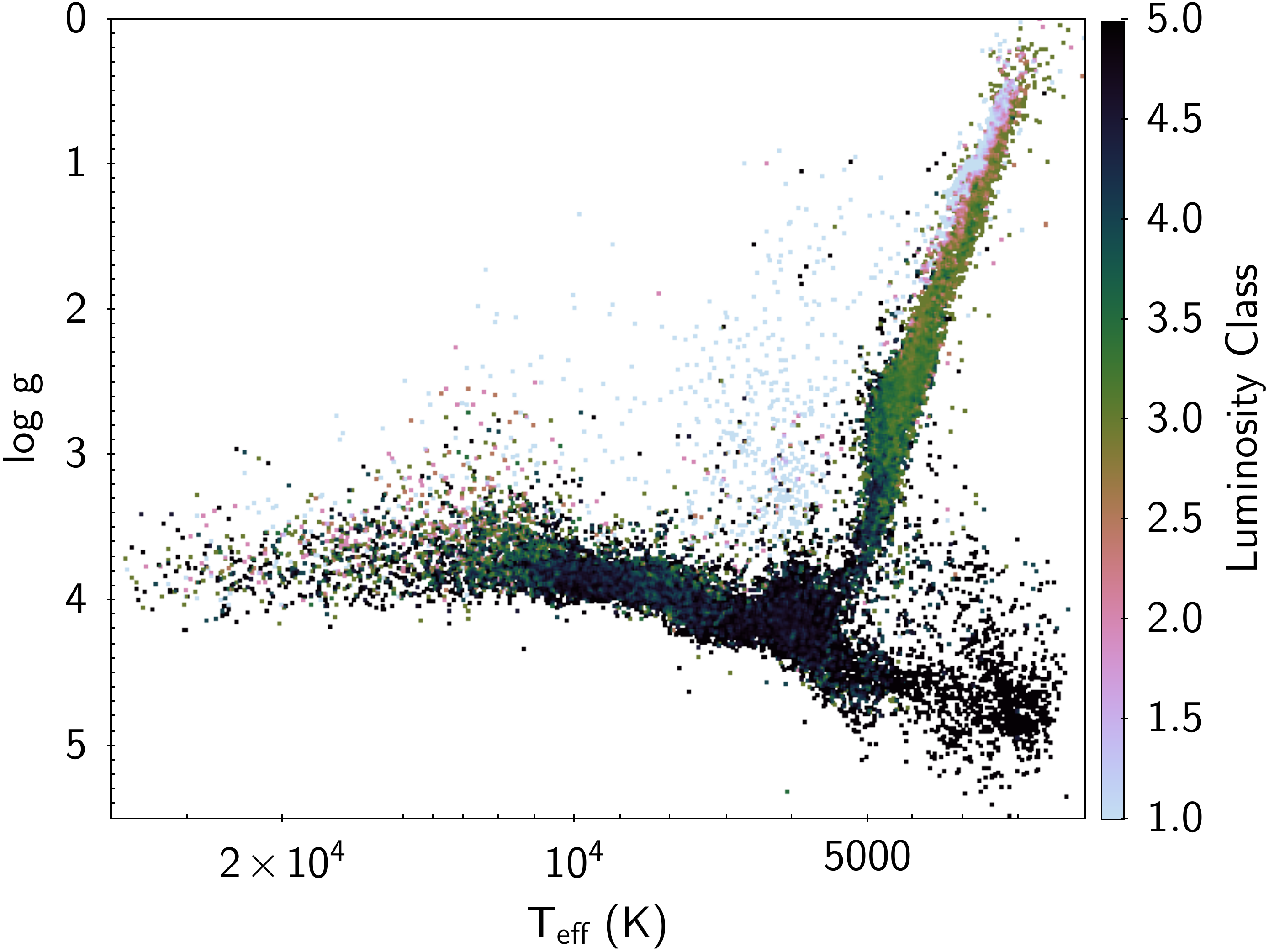}
\caption{Left: Comparison between spectral types in literature and the \teff\ measured for the stars in the sample, color coded by the luminosity class. Yellow circles show the sources with spectral types measured directly from the APOGEE spectra, from \citet{roman-lopes2019,roman-lopes2020,ramirez-preciado2020}. Spectral types are formatted to range as O=0-10, B=10-20, A=20-30, etc. Right: distribution of \teff\ and \logg\ color coded by the literature luminosity class. Note that the concentration of the supergiants at \teff$\sim$6000 K and \logg=3.5 is primarily from the Magellanic clouds (Section \ref{sec:mc}). \label{fig:sptmatch}}
\end{figure*}

\begin{figure*} 
\epsscale{1.1}
\plotone{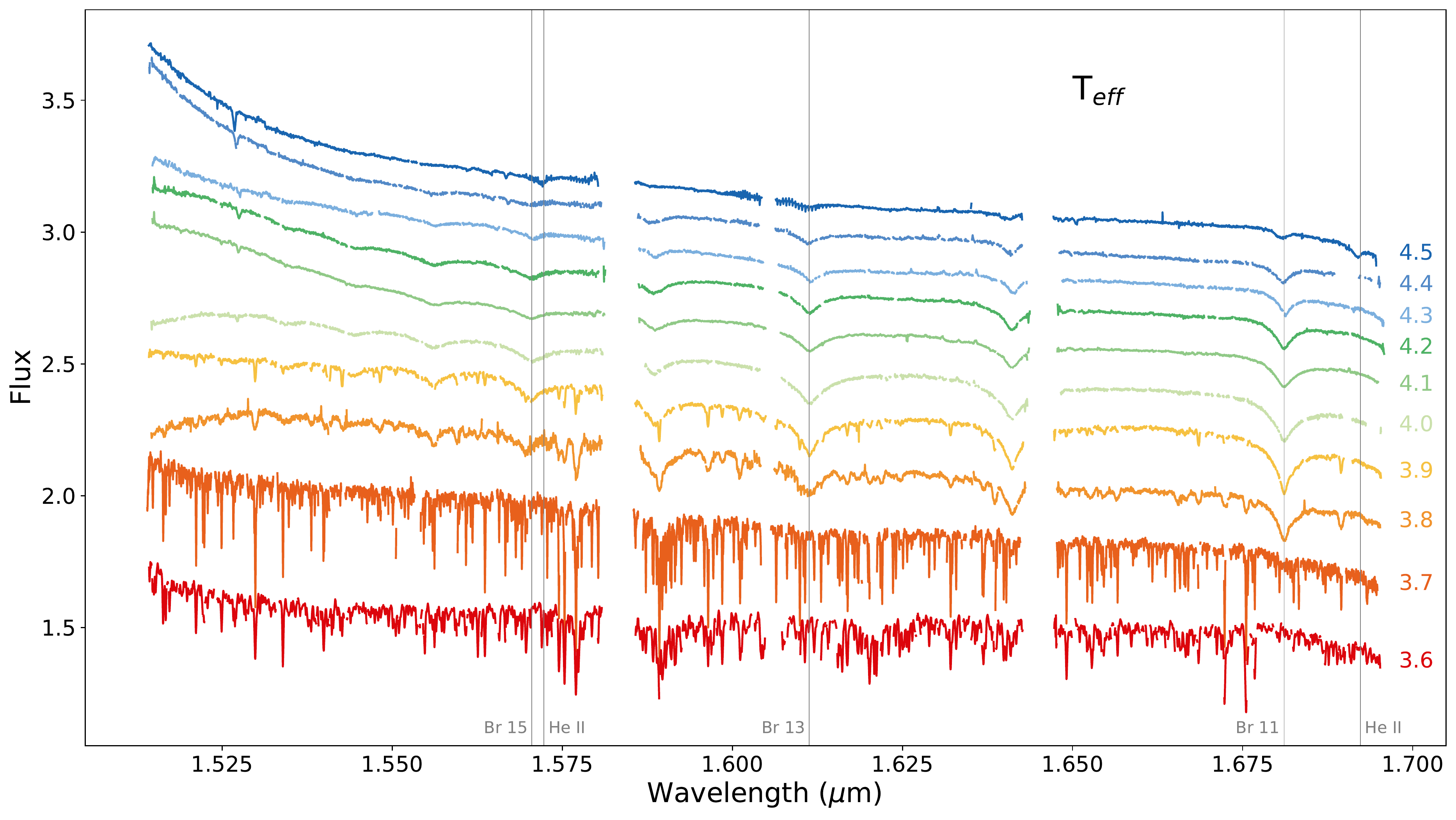}
\plotone{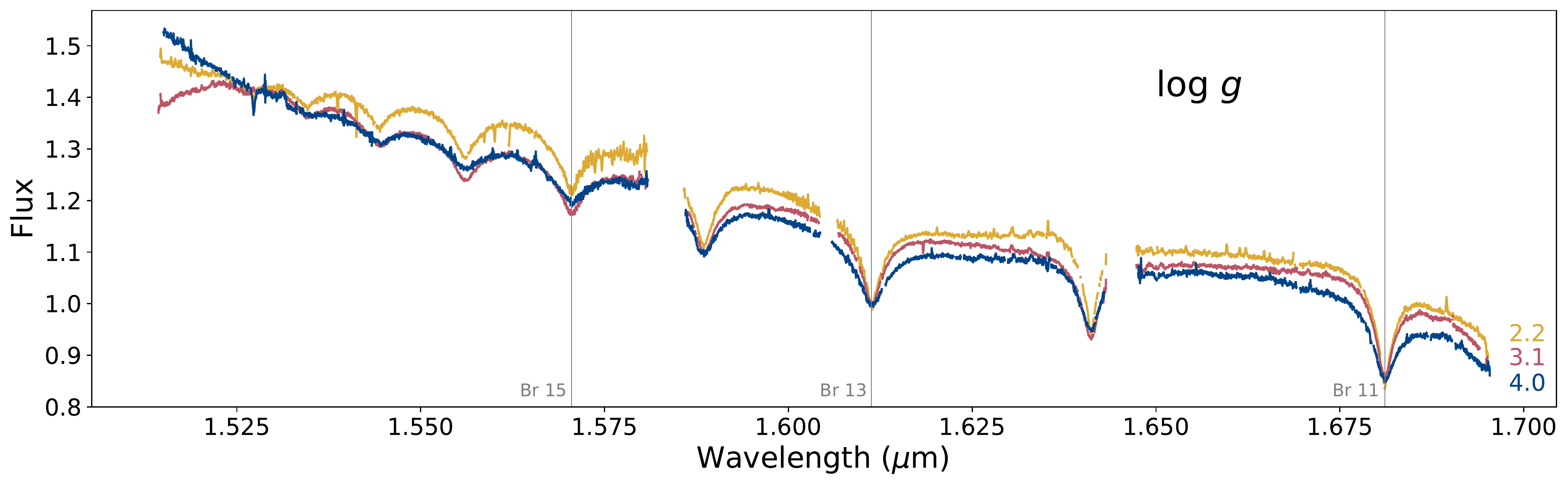}
\caption{Top: Example APOGEE spectra for sources with \logteff\ ranging from 4.5 (blue) to 3.6 (red), with a step of 0.1 dex. The spectra are arbitrarily scaled in flux. Some of the temperature sensitive lines that can be used to infer parameters of hot stars are shown in gray, such Brackett H lines and He II lines. The gaps in the spectra correspond to the chip gap. Bottom: Spectra with similar \logteff$\sim$4.1, but with three different \logg.\label{fig:teffgrid}}
\end{figure*}

\begin{figure*} 
\epsscale{1.1}
\plottwo{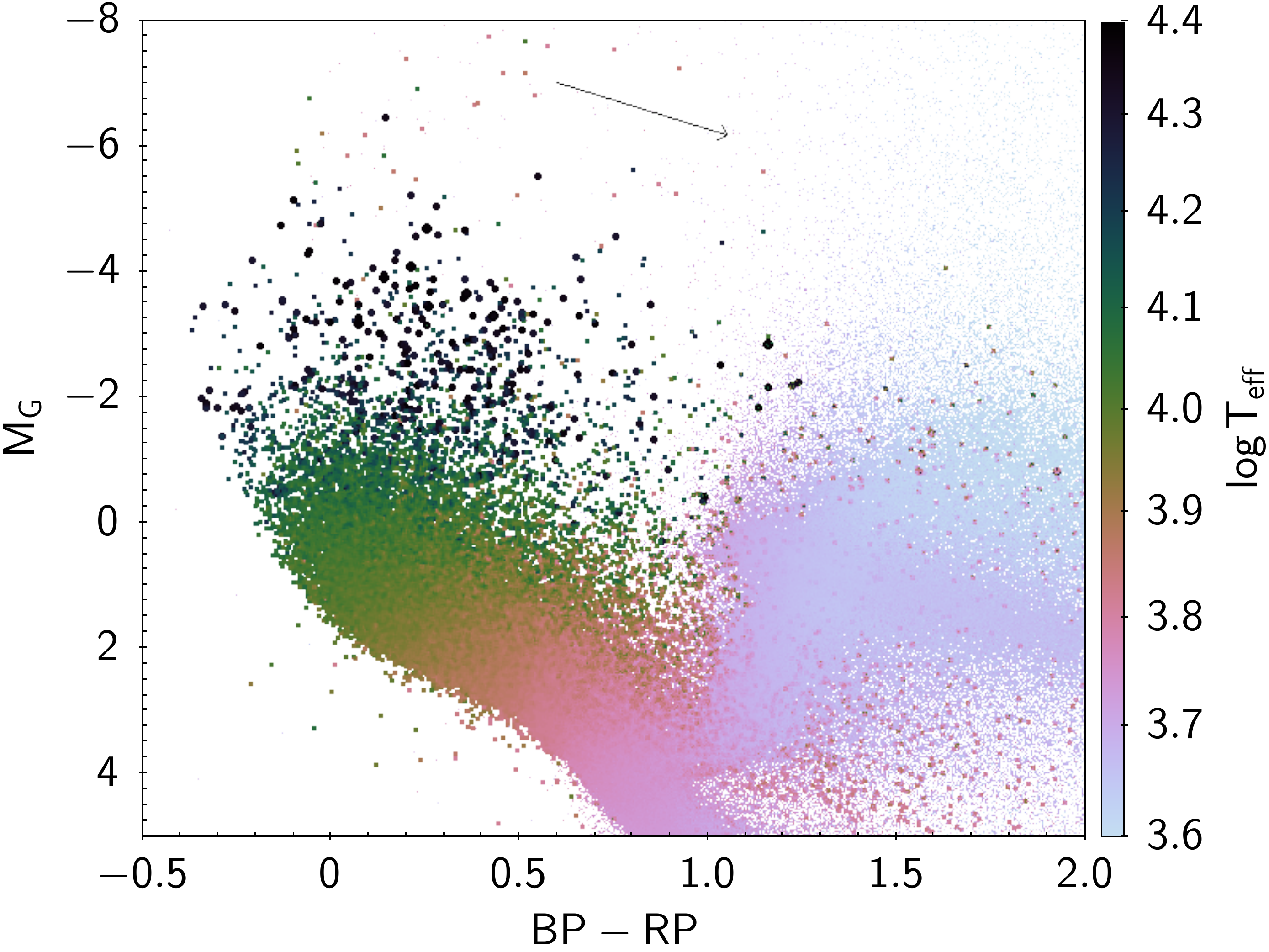}{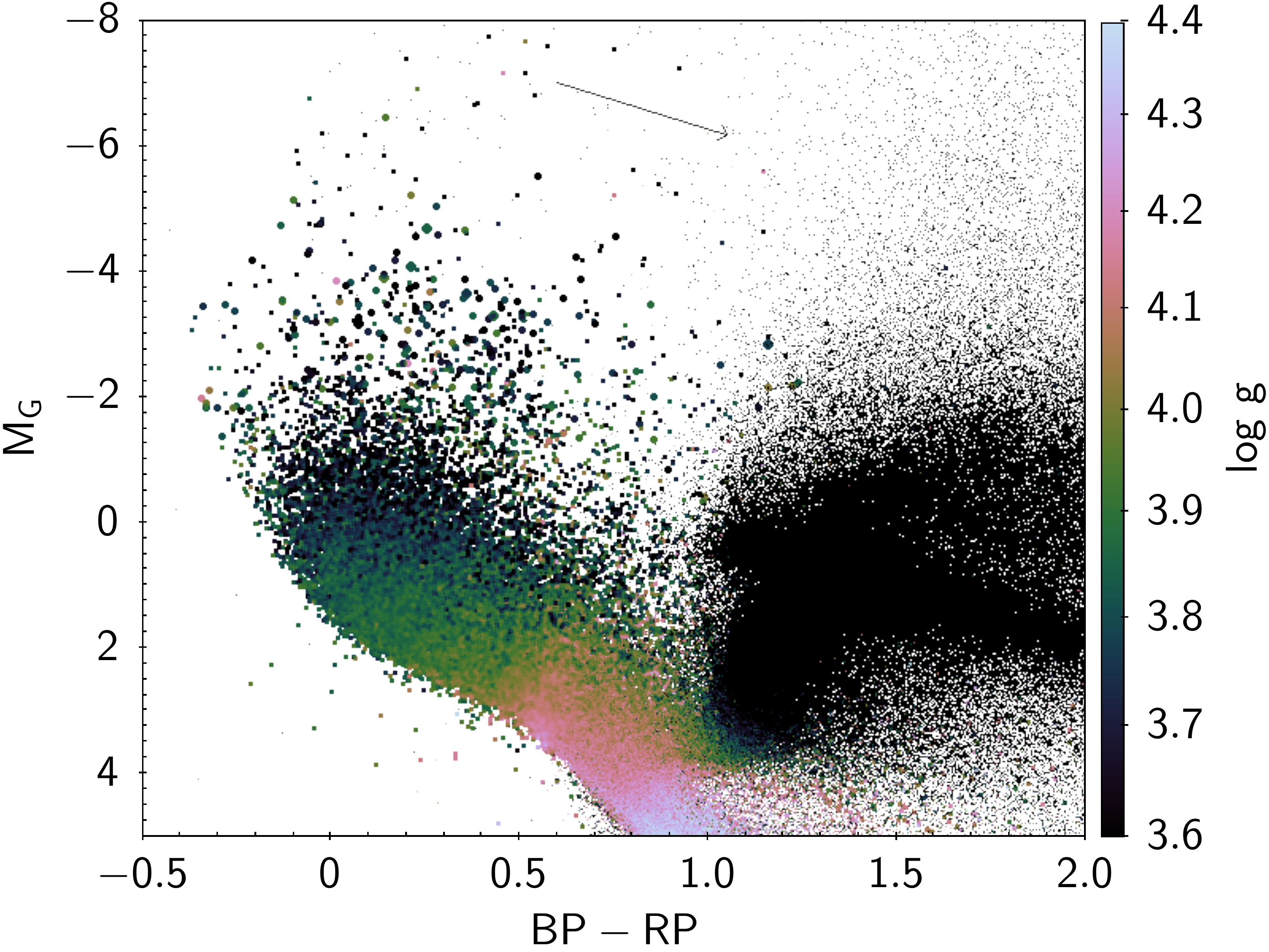}
\caption{HR diagram of the APOGEE sample of higher mass stars, color coded by the predicted \teff\ (left) and \logg\ (right). Black arrows show the reddening vector corresponding to 1 $A_V$. \label{fig:hr}}
\end{figure*}

In general, as APOGEE Net II matches the performance of APOGEE Net I for low mass stars; the primary improvement in the pipeline is in its new ability to generalize the parameters of high mass stars. Indeed, the derived \teff\ are strongly correlated to the previously measured spectral types for the same stars (Figure \ref{fig:sptmatch}), particularly among the main sequence stars. There is some scatter; for the high mass stars this scatter is in part driven by multiplicity. As almost all O stars tend to be multiples \citep{preibisch2001,duchene2013}, it is not uncommon for spectral features of both stars to be detected in the same spectrum. Although spectral types can contain information on both stars (e.g., O5+O9.5), such information is difficult to encode, as such, a comparison is done for only one star in the pair, but the derived \teff\ may favor a second star.

Different wavelength regimes may also be more or less sensitive to the presence of a companion, as such, optical spectra, from which most spectral types have been derived may not offer a perfect match to the H band APOGEE spectra. As such, the correlation between spectral type and the measured \teff\ is particularly strong among B stars that had spectral types derived directly from the APOGEE spectra \citep{ramirez-preciado2020} rather than those measured from other datasets.

The hot main sequence stars (\teff$>10^4$ K, \logg$>3.7$) are relatively numerous in the sample and they tend to have a set \logg\ as a function of their \teff\, and their \teff\ (and the spectral features that correspond to it) vary smoothly as a function of the mass of the star. However, this is not the case for blue giants and blue supergiants. Only a few hundred of these stars with luminosity classes from the literature have been observed. There is also a large gap in \logg\ between the blue giants and blue supergiants (e.g., Figure \ref{fig:grid}, difference between red and green lines); this gap is difficult for a CNN to fully reproduce, especially given a limited sample size owing to the rarity of such objects, resulting in \logg\ of supergiants being overestimated and placing them closer to \logg\ distribution of other stars. Finally, luminosity classes at \teff$<8,000$ K become less precisely defined relative to \logg\ (e.g., Figure \ref{fig:grid}). Combined, these effects make it difficult to achieve optimal performance in extracting stellar parameters of these type of stars. The mismatch in \logg\ for the supergiants between the training labels and those predicted by the model is partially apparent in Figure \ref{fig:onetoone}, but these objects are extremely rare. 

However, we note that qualitatively, examining the distribution of hot stars in the right panel of Figure \ref{fig:sptmatch} color coded by their luminosity types does show a preference for assigning more luminous types to the sources with lower \logg\ at a given \teff. Thus, although their \logg\ may be overestimated, it can nonetheless be used to separate stars of different luminosity classes.

Examining the spectra directly, they can be sorted based on their \teff\ measurements into a sequence (Figure \ref{fig:teffgrid}, top). While hot stars lack the number of spectral features that are present in cooler stars, equivalent widths of H lines that fall into the APOGEE spectrum, combined with various atomic features (for \teff$<$10,000 K), as well as He II absorption (for \teff$>25,000$ K), are effective at determining \teff. Similarly, surface gravity broadening is imprinted on the H lines: while \logg\ is more difficult to measure, dwarfs do appear to have somewhat wider lines than the giants (Figure \ref{fig:teffgrid}, bottom).

Another method of evaluation is via HR diagram. Figure \ref{fig:hr} shows that the bluer stars are preferentially hotter as well, and that the more luminous stars tend to have lower \logg. It should be noted that nearby OB stars are too bright to be targeted for the spectroscopic observations with the current targeting strategy, as such hot stars tend to be more distant and more susceptible to be reddened due to extinction. Indeed, the gradient of constant \teff\ or \logg\ color coded on the HR diagram does appear to follow the reddening law, and it is possible to deredden the hottest OB stars (\logteff$>4.2$) to the tip of the main sequence.

\begin{figure*} 
\epsscale{1.1}
\plotone{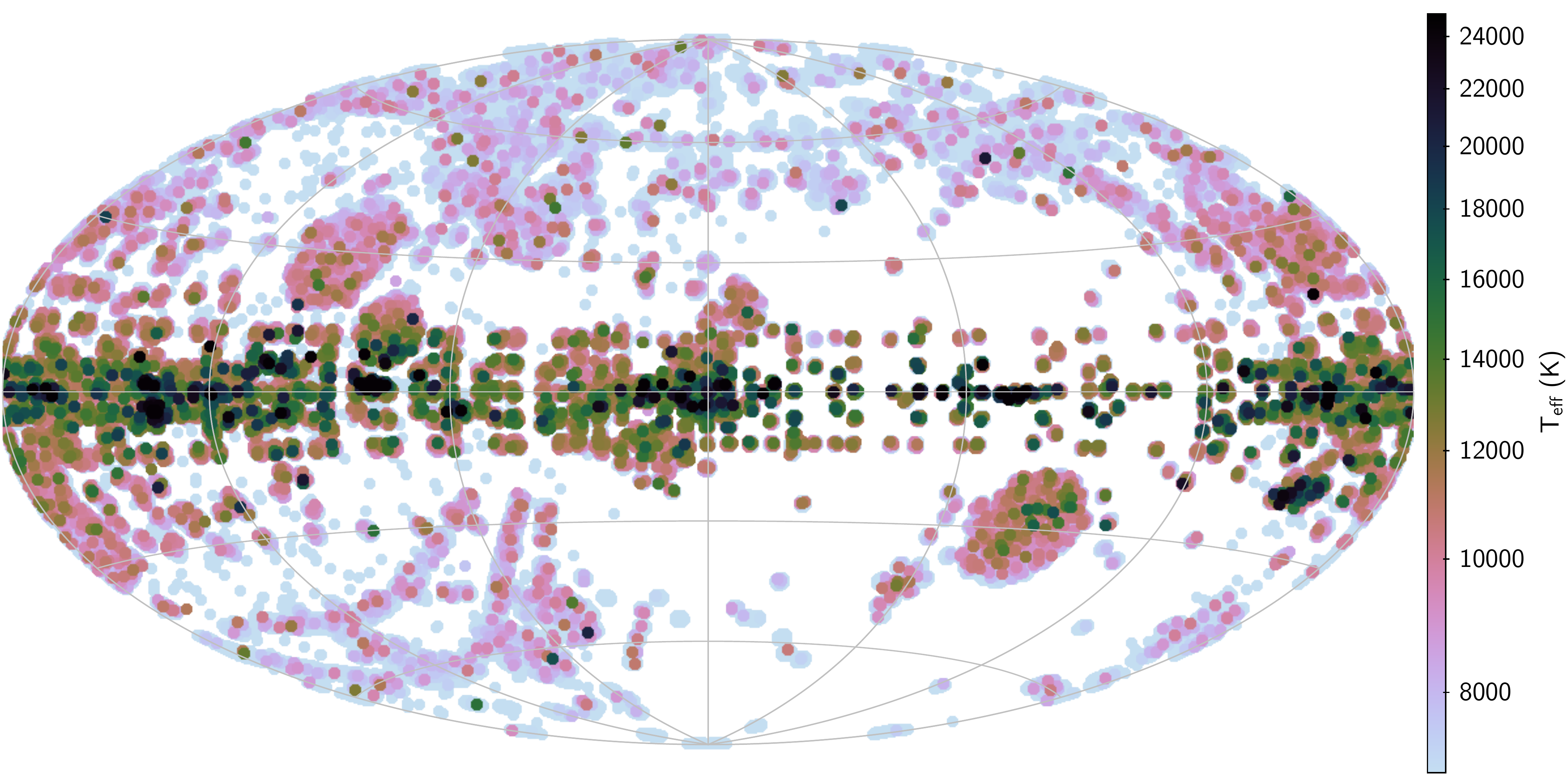}
\plotone{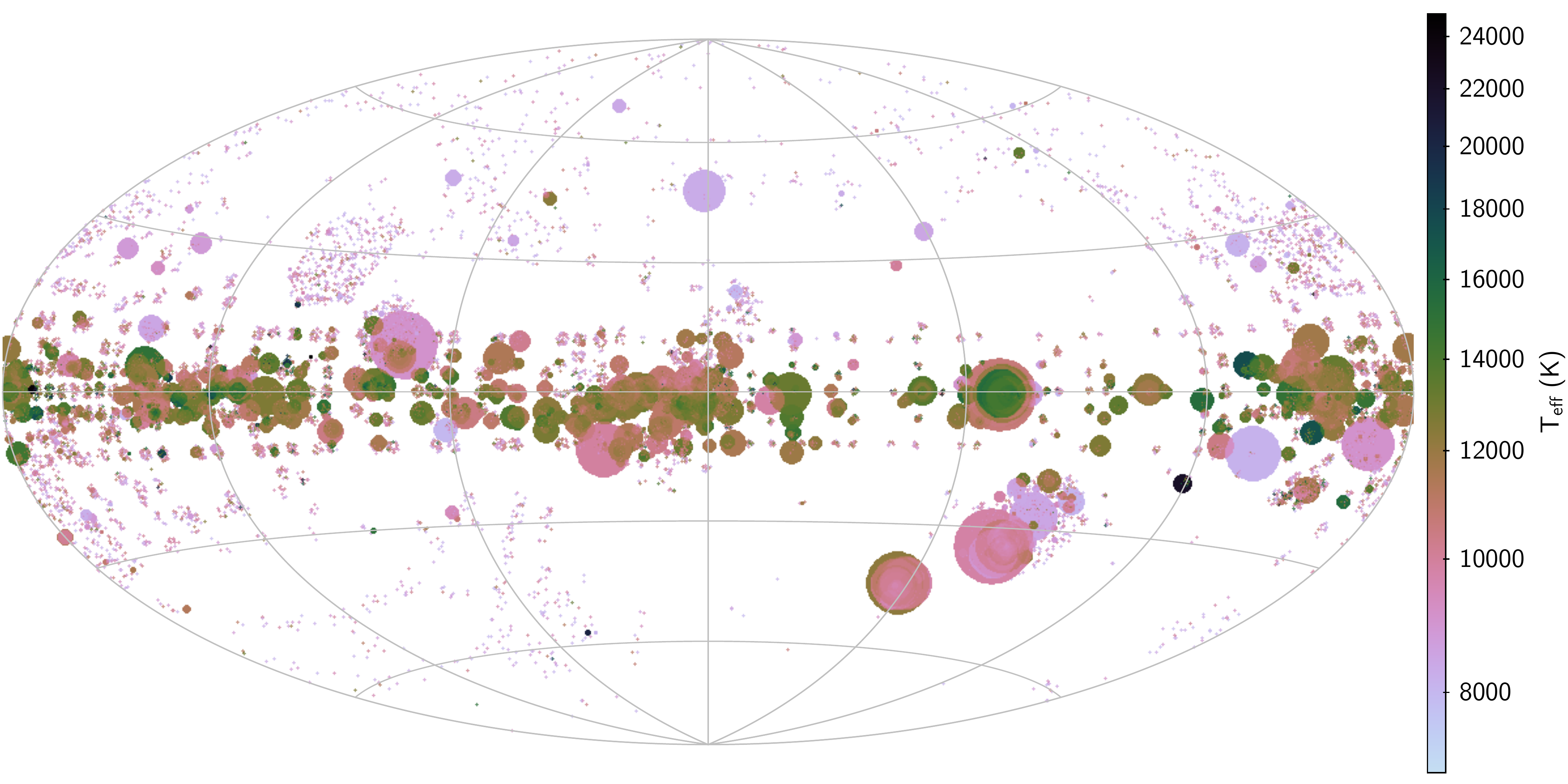}
\caption{Top: Distribution of sources in the APOGEE DR17 data in galactic coordinates, color coded by the maximum \teff\ of a source along a line of sight. Note that sources with \teff$>$10,000 K tend to be concentrated in the plane, and sources with \teff$>$20,000 K tend to primarily trace young star forming regions. Bottom: distribution of the sources, but limited to only sources with \teff$>$8,000 K, with the symbol size representative of \logg$<$3.5, to highlight the position of blue and yellow giants and supergiants. \label{fig:obsky}}
\end{figure*}

The final method of evaluation of the accuracy in the determination of \teff\ and \logg\ of high mass stars with respect to the other sources is through examining the distribution of the sources within the Galaxy. As high mass stars have short lifetimes, they are not expected to be located far from their regions of birth. Indeed, the hottest stars (\teff$>20,000$ K) have a very clumpy spatial distribution that almost exclusively trace known star forming regions. Similarly, somewhat cooler stars ($10,000<$\teff$<20,000$ K) are located primarily within the Galactic disk, with the disk scale height increasing with decreasing \teff\ (Figure \ref{fig:obsky}), as the disk scale height also depends on the age of the stellar population tracer being employed.

Similar distribution of sources is observed among the blue and yellow supergiants. Although their \teff\ is lower than for the main sequence counterparts of the same mass, selecting sources with, e.g., \teff$>$8,000 K and \logg$<$3.5 also preferentially traces younger parts of the Galaxy.

\subsection{Magellanic clouds}\label{sec:mc}

The APOGEE-2 survey has primarily observed the stars of the Milky Way, as such, APOGEE Net has been trained in the Milky Way stars almost exclusively - however, there have been several dedicated programs targeting stars in the Magellanic clouds \citep[MCs,][]{santana2021,zasowski2017,nidever2020}). These galaxies are significantly more metal-poor \citep{nidever2020,hasselquist2021}, as such, the unfamiliar chemistry and other features unique to MCs may skew some of the weights within the APOGEE Net model, particularly as abundances also affect \teff\ and \logg\ due to various elements playing a significant role in the energy transport in the outer structure of a star. As such, these extreme conditions result in particularly informative region to evaluate the performance of the network.

\begin{figure*} 
\epsscale{1.1}
		\gridline{\fig{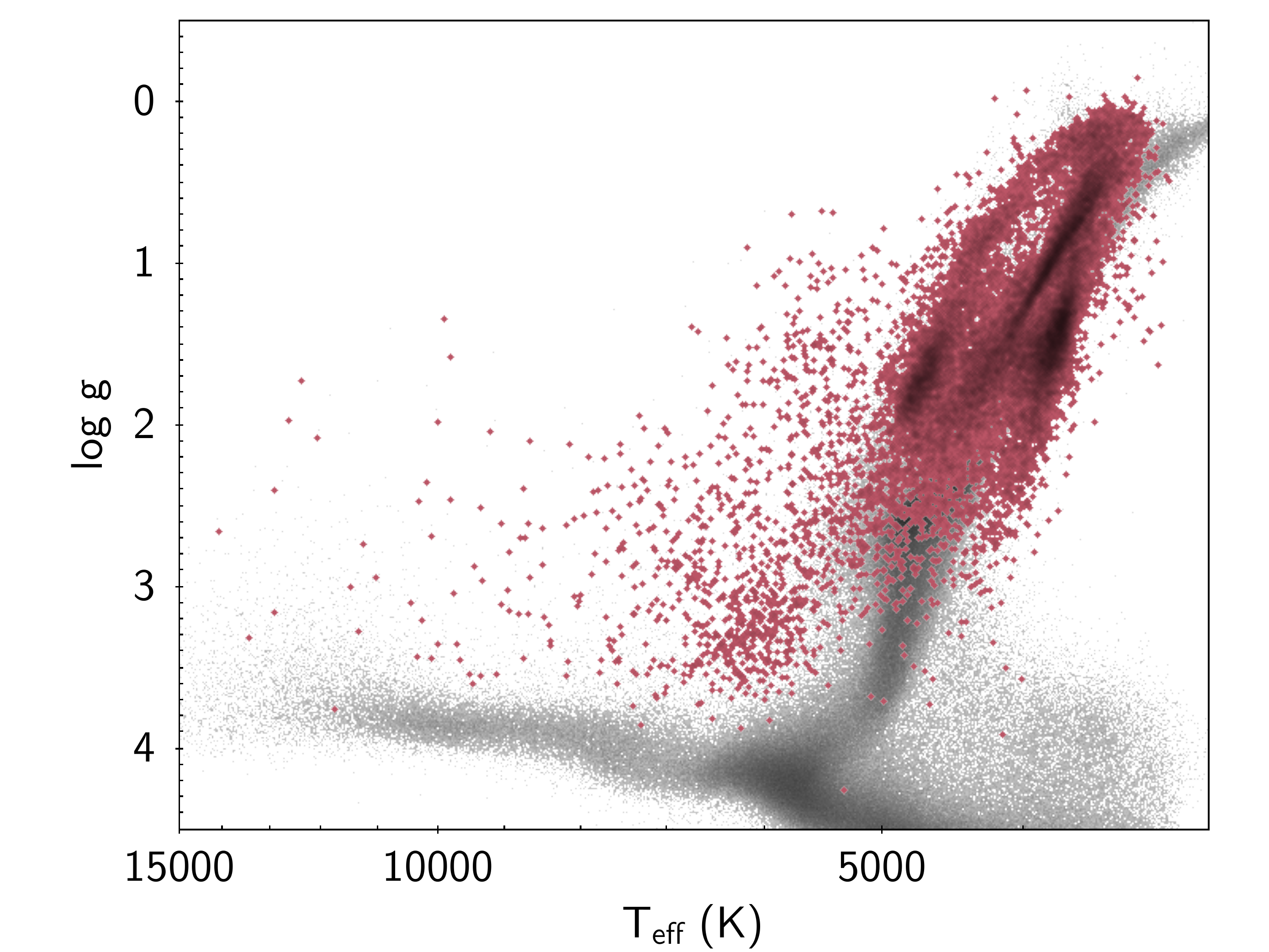}{0.31\textwidth}{}
                  \fig{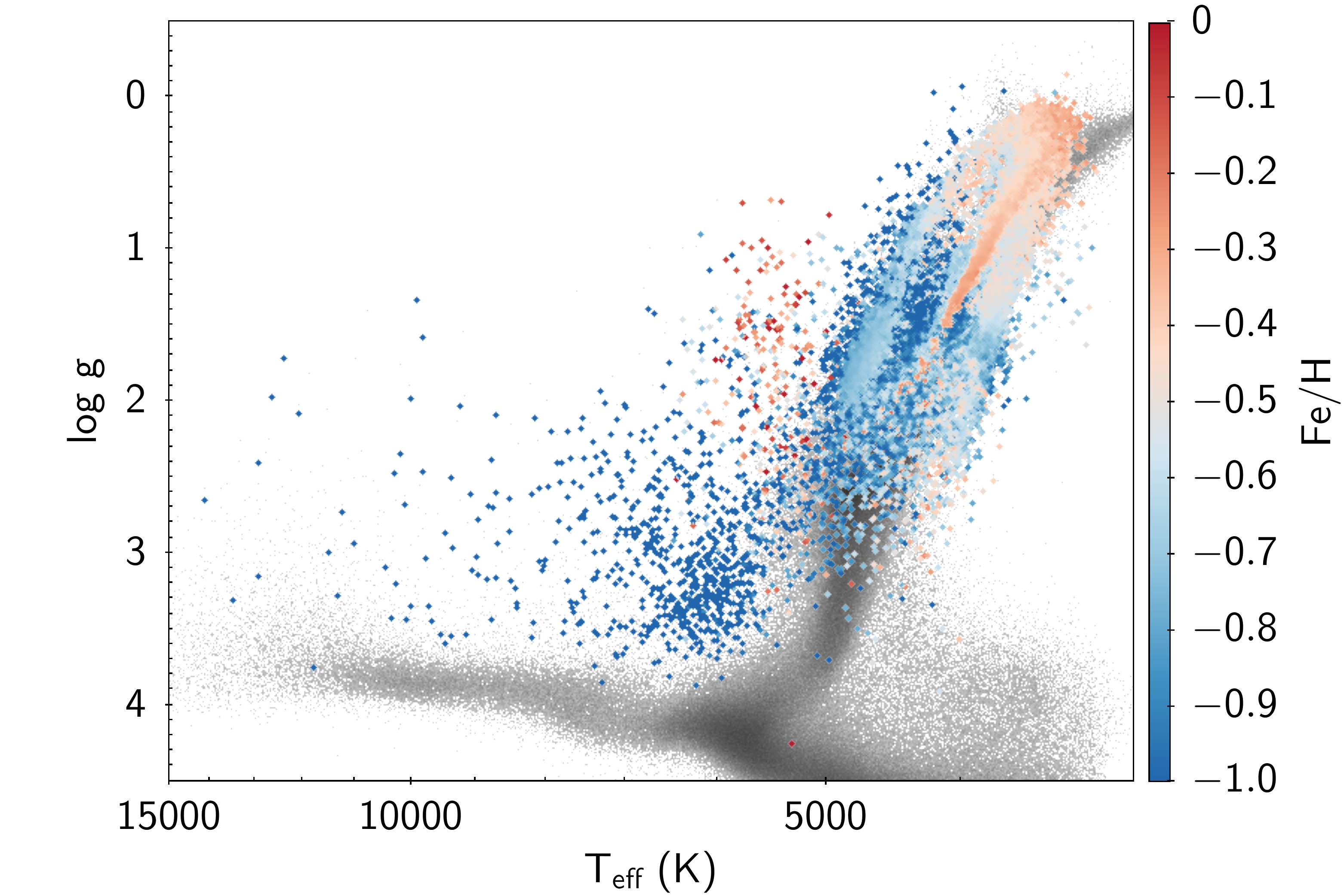}{0.34\textwidth}{}
                  \fig{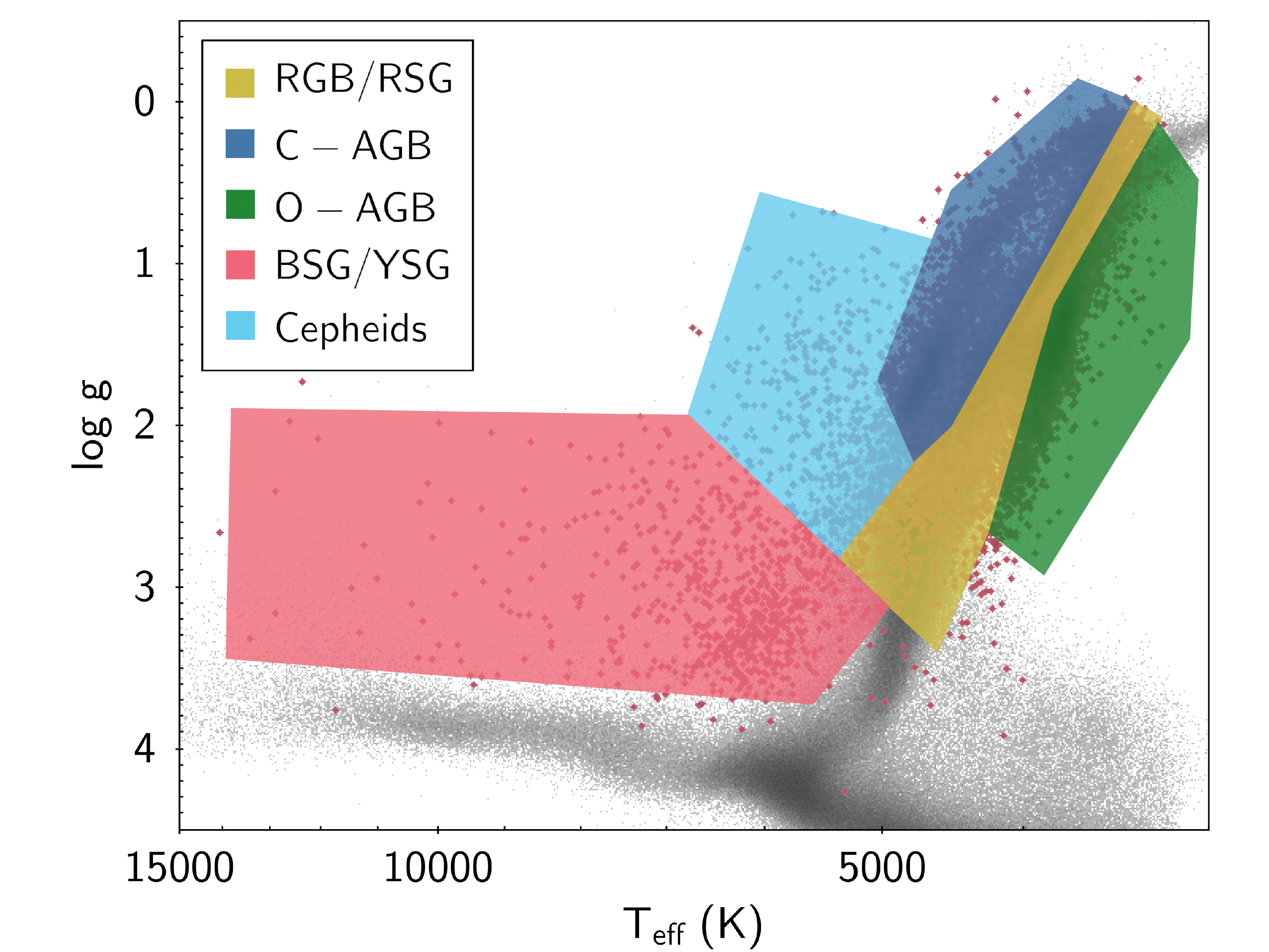}{0.31\textwidth}{}
        } \vspace{-0.8cm}
		\gridline{\fig{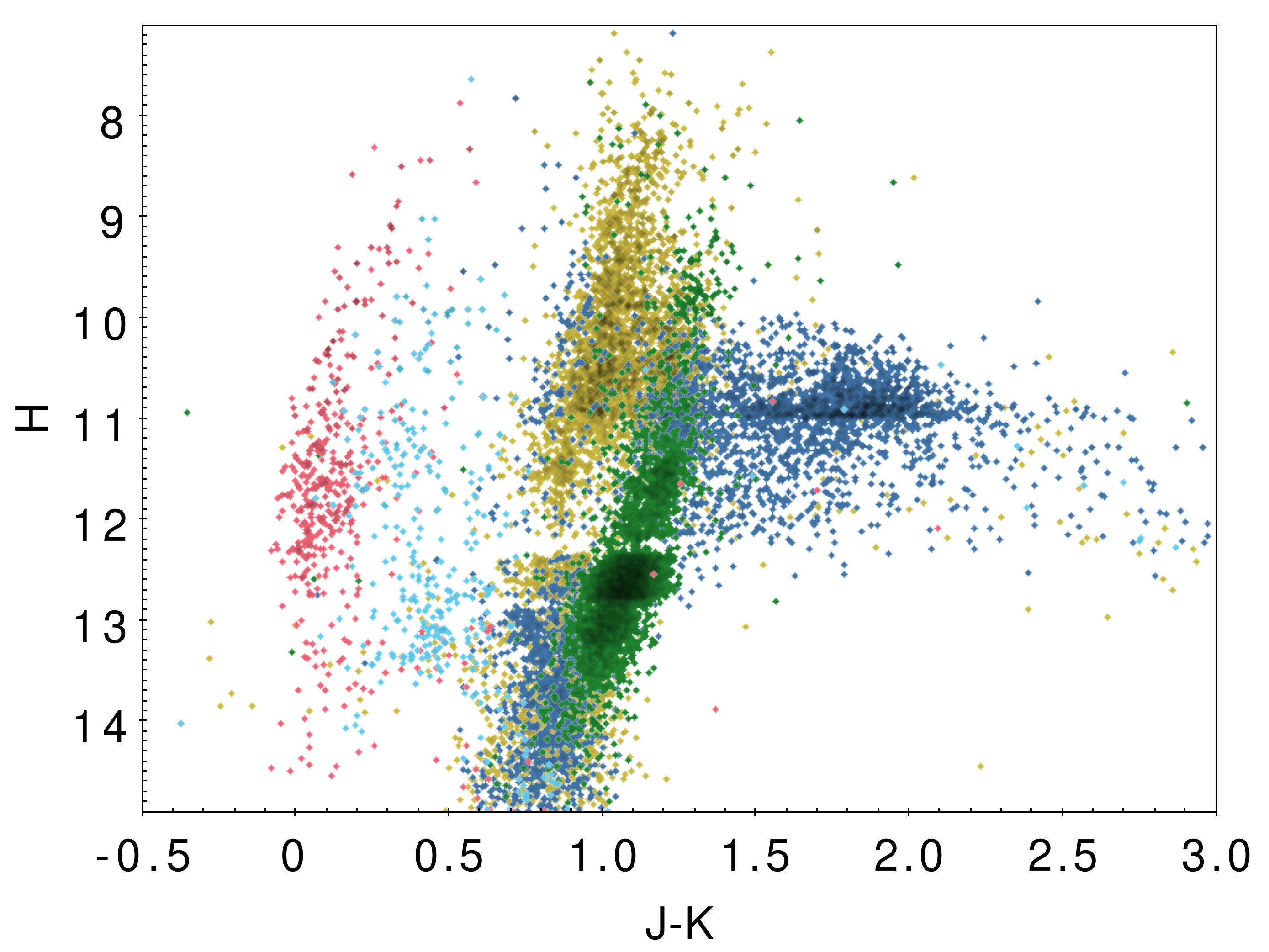}{0.33\textwidth}{}
                  \fig{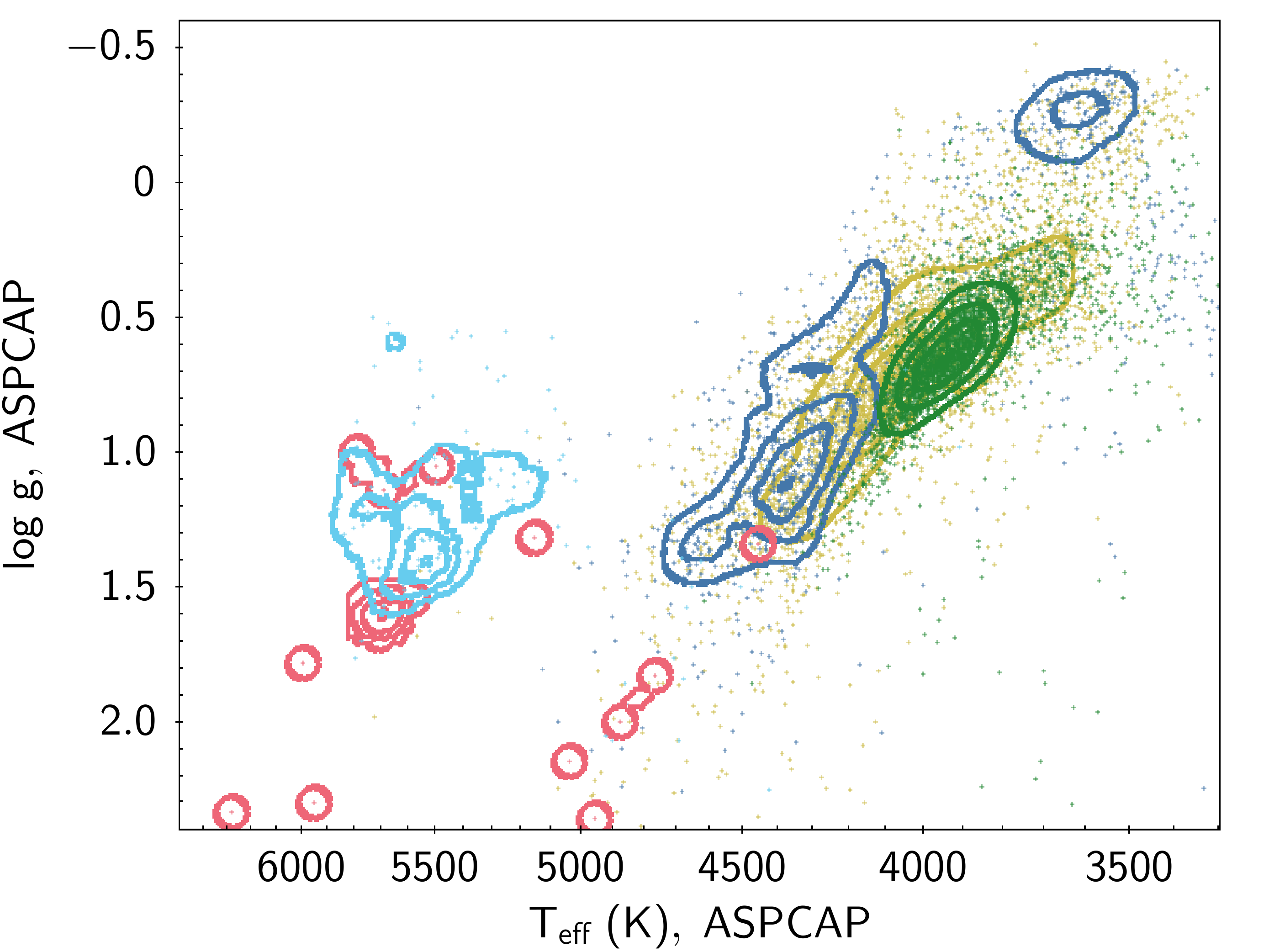}{0.33\textwidth}{}
        } \vspace{-0.8cm}
		\gridline{\fig{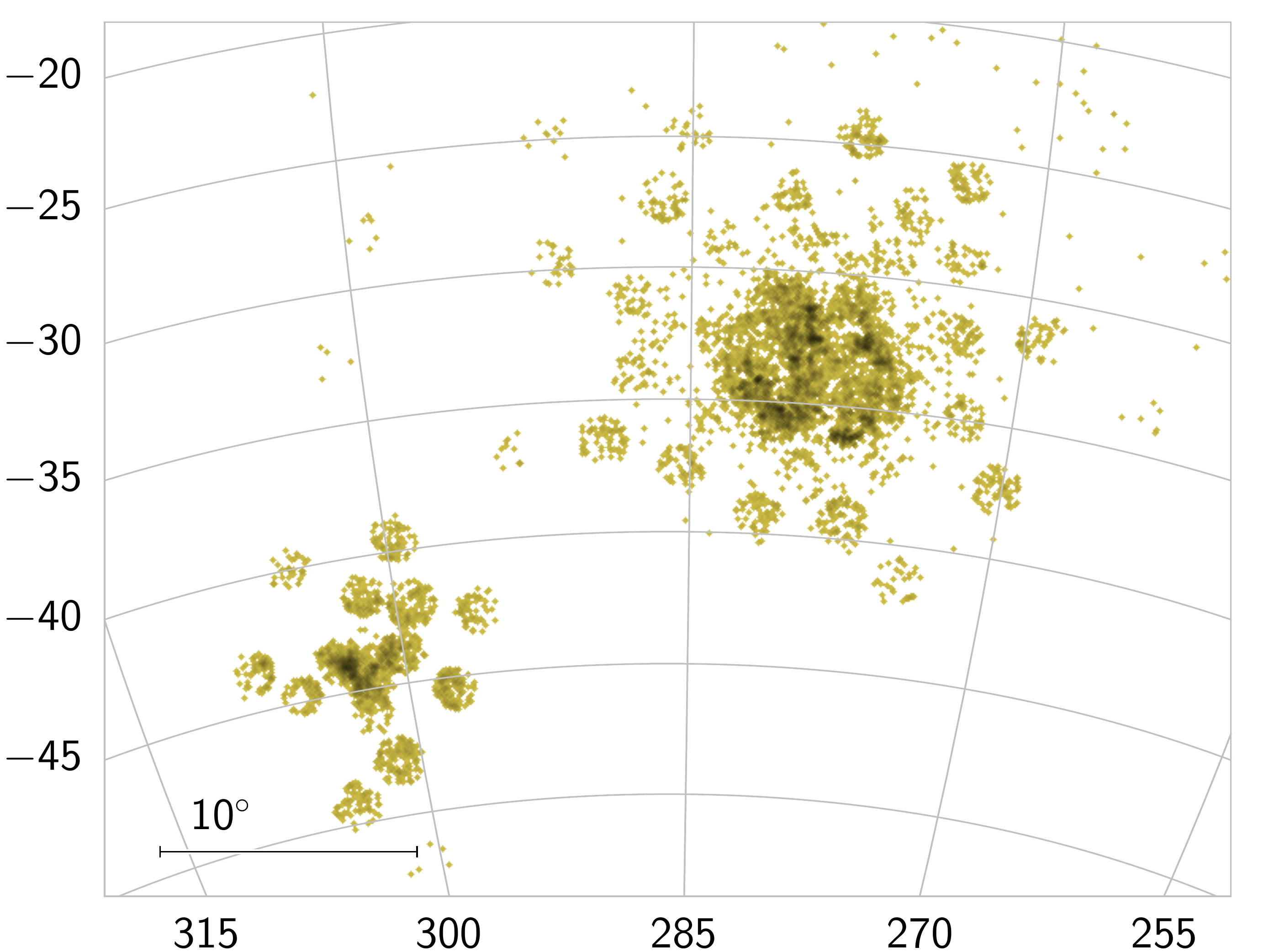}{0.33\textwidth}{}
                  \fig{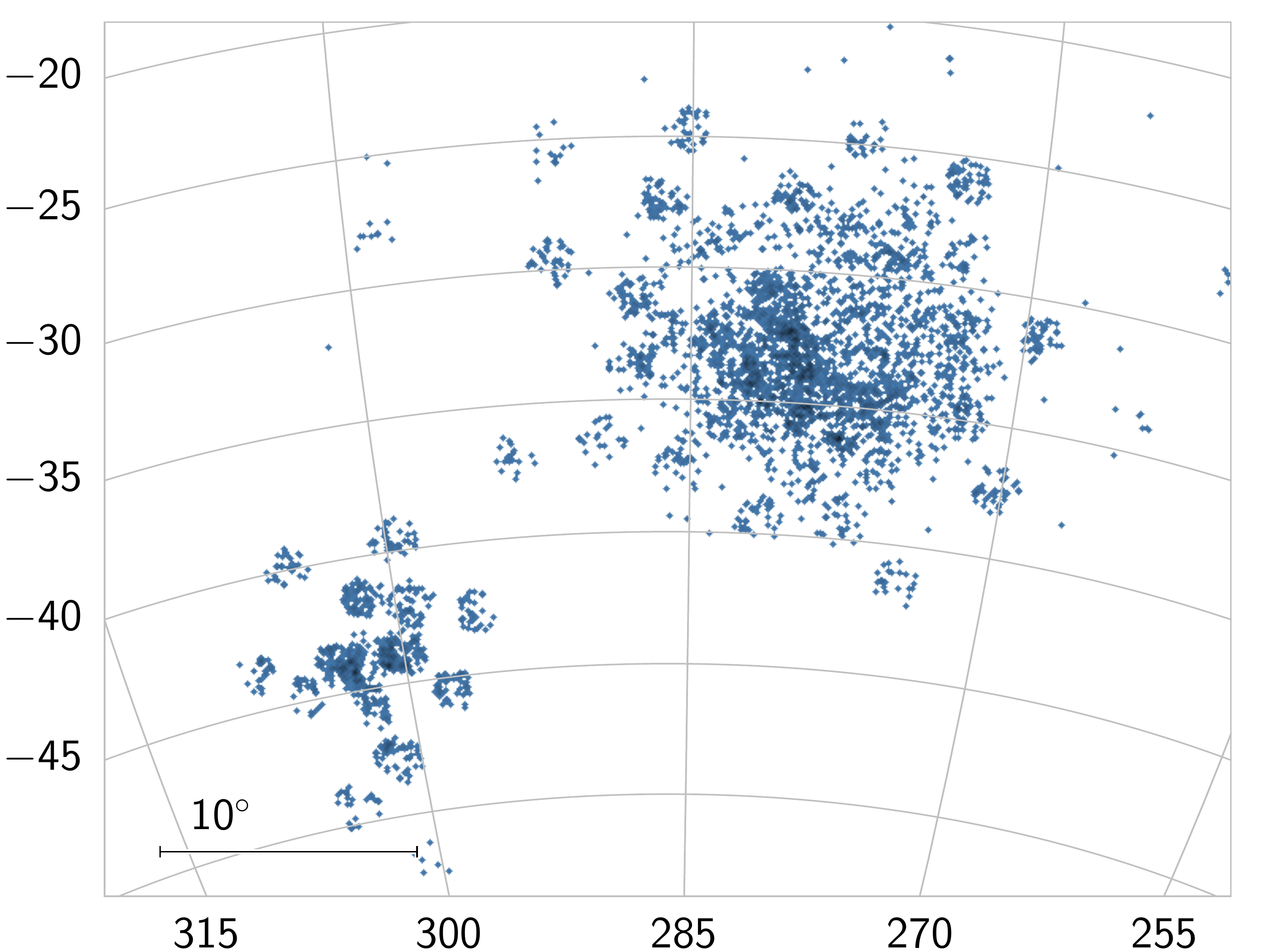}{0.33\textwidth}{}
                  \fig{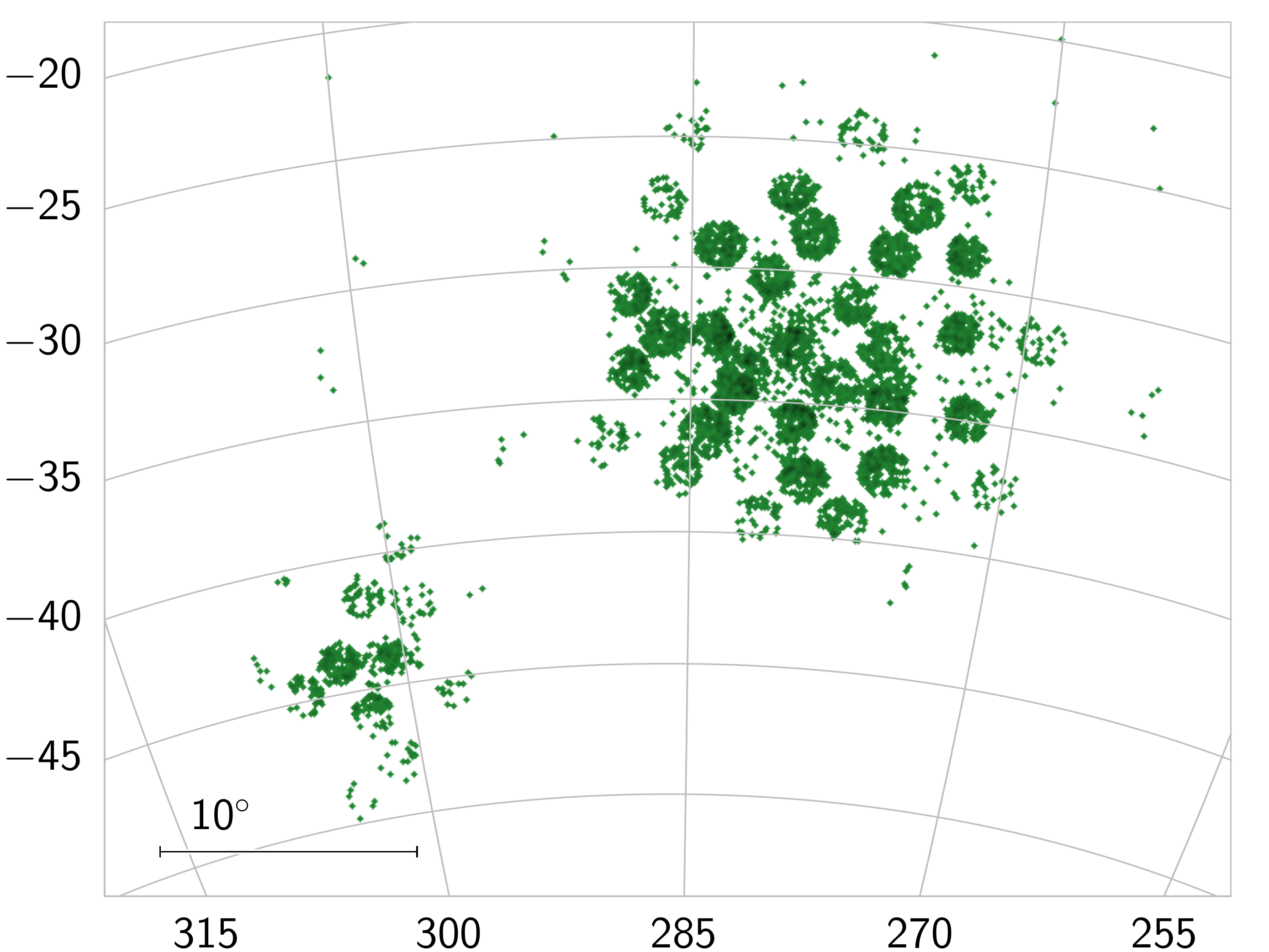}{0.33\textwidth}{}
        } \vspace{-0.8cm}
		\gridline{
                  \fig{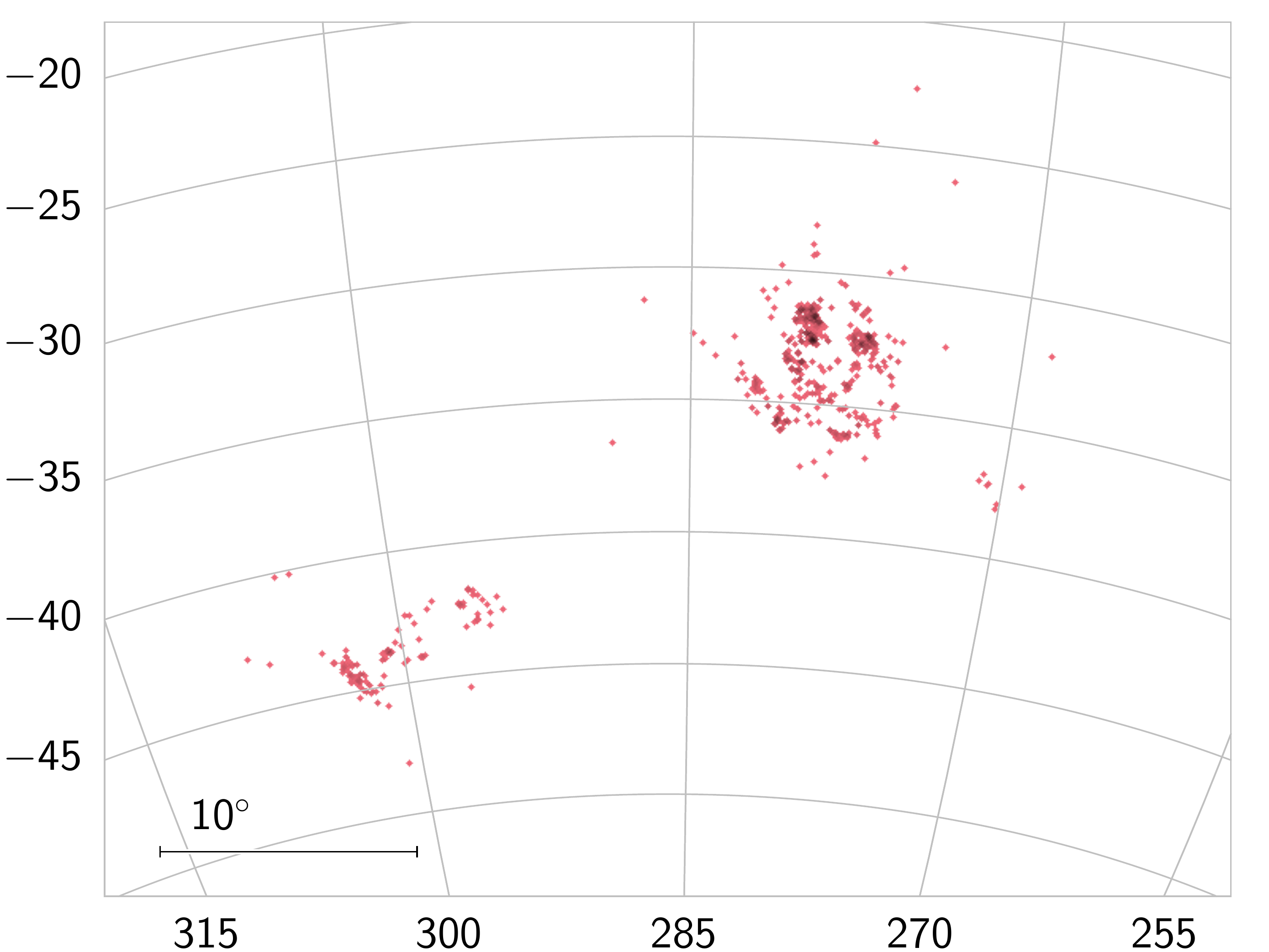}{0.33\textwidth}{}
                  \fig{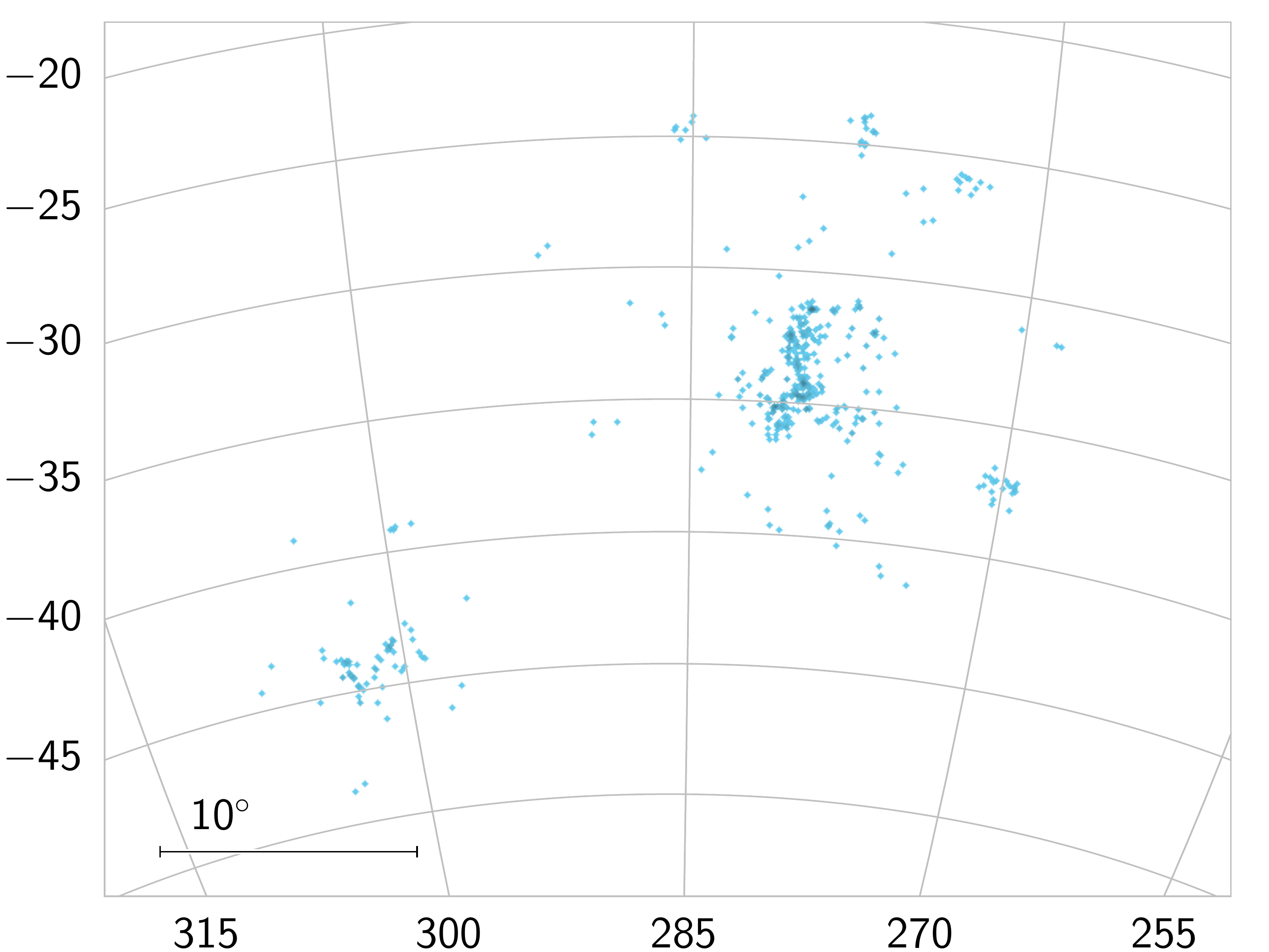}{0.33\textwidth}{}
        } \vspace{-0.8cm}
\caption{Stars observed by APOGEE towards the Magellanic clouds. Top left: Distribution of \teff\ and \logg\ of the likely members (red), superimposed on the Milky Way stars in grayscale background. Top middle: Same as before, only with stars color-coded by their estimated [Fe/H]. Top right: Same as before, but separating the stars into five categories based on the parameter space: red giant branch/red supergiant (RGB/RSG), carbon-rich AGB stars (C-AGB), oxygen-rich AGB stars (O-AGB), blue and yellow supergiants (BSG/YSG), and the Cepheid variables. Second row left: 2MASS color magnitude diagram of the stars that fall into this parameter space. Second row right: ASPCAP parameters for the stars in this parameter space. Although three sequences of RGB/AGB stars overlap, they are nonetheless distinct from one another in this dataset as well. Other panels show the spatial distribution of the stars in these categories. \label{fig:mc}}
\end{figure*}

We select the MC members based on the parallax ($<$0.05 mas) and radial velocity ($>$100 km s$^{-1}$). Examining the distribution of \teff\ and \logg\ of these stars does show some surprising features: the red giant branch appears to be trifurcated, in a manner that is not seen in the APOGEE sample of the Milky Way (Figure \ref{fig:mc}). However, there are physical interpretations for this segmentation: the stars within each overdensity in the spectroscopic parameter space do trace different spatial regions in the MCs, and some of them have a particular targeting flag. Furthermore, the relative placement of these segments is mostly consistent with the parameters derived from ASPCAP. However, in ASPCAP the sequences are somewhat closer together, partially overlapping, producing a more continuous gradient. This could be due to the fact that in producing a fit, ASPCAP is considering all stellar parameters (including \teff, \logg, and all of the elemental abundances) independent variables, and \teff\ could be skewed by the influence of opacity contributed by the C and O abundances.

Along the red giant branch (RGB), the middle of the three overdensities (Figure \ref{fig:mc}, yellow) are the common RGB stars and red supergiants (RSG); the location of this branch is consistent with what is observed in the Milky Way. 

The hotter branch (Figure \ref{fig:mc}, dark blue) is somewhat more peculiar. Though there are stars in the Milky Way that do fall into that parameter space, they do so solely due to being regular RGB stars with low metallicity, there is not a concentration of them at \teff$\sim$5000 K \logg$\sim$1.8 as there is in MCs. Their placement on 2MASS color-magnitude diagrams suggests that they are likely to be carbon rich AGB-C stars \citep{boyer2011,nidever2020}.

The cooler branch (Figure \ref{fig:mc}, green) is particularly unusual, as there are no counterparts to these stars in the APOGEE sample in the Milky Way. However, all of these sources share the same targeting flag of APOGEE2\_WASH\_GIANT, which is a classification that pre-dates the release of Gaia DR2, performed using Washington filter system to separate likely dwarfs and giants. Though sources with WASH\_GIANT flag have been observed from both northern and southern hemisphere, the peculiar concentration is limited only among the members of the Magellanic clouds. This clump also matches well to the variable stars identified by OGLE, usually as long period variables with the subtype of small amplitude red giants \citep{soszynski2009,soszynski2011}. The placement of these stars along the color-magnitude diagram suggests they are oxygen rich AGB-O stars \citep{boyer2011}. 

Stars with \teff$>$6000 K and \logg$>2$ (Figure \ref{fig:mc}, red) appear to be young blue and yellow supergiants, and they appear to be spatially concentrated in the regions of active star formation.

Sources with \logg$\sim$2 and \teff$\sim$6000 K (Figure \ref{fig:mc}, light blue) have different metallicity from the supergiants that is closer towards solar. Most of these sources have been specifically targeted for being Cepheid variables. 

\section{Conclusions}

We present stellar parameters from the APOGEE spectra derived via a neural network, APOGEE Net. This is the first pipeline for analyzing these data that is capable of extracting stellar parameters for all stars, regardless of their temperature (from $\sim$3,000 K to $\gtrsim$50,000 K, which surpasses the hot limit of 20,000 K reported by ASPCAP in DR17) or surface gravity (from $\sim$0 to $\sim$5 dex), in a self-consistent manner. This includes pre-main sequence stars, main sequence dwarfs, red giants, massive main sequence stars, and blue supergiants.

These parameters do have some dependence on the trained neural network model: although this is less of an issue for common objects, the rarer classes of objects may not fully reach the precise parameter space these objects are supposed to inhabit, e.g., blue supergiants may have their \logg\ somewhat overestimated (closer to the main sequence) than the parameters in the training set may imply. Nonetheless, the network does appear to place stars along a representative sequence in both \teff\ and \logg\, as such these types of stars can nonetheless be identified and selected.

In addition to the stars observed in the Milky Way, APOGEE Net appears to achieve adequate performance in regions with less familiar (to the neural network) chemistry, such as the Magellanic clouds. Although it does produce some surprising features in the \teff\ \& \logg\ parameter space (which are also present in other pipelines, such as ASPCAP, but to a lesser degree, possibly due to allowing for independent parameters for C \& O in the fitting process), these features do appear to be physically motivated, identifying distinct classes of objects.

APOGEE DR17 is the final data release produced under SDSS-IV. The next generation of the survey, SDSS-V, intends to obtain spectra of $>6$ million stars, providing a much more homogeneous spectroscopic coverage of the Milky Way \citep{kollmeier2017}, including young stars along the Galactic disk, both low mass and high mass. APOGEE Net and its subsequent iterations provide the means to uniformly infer stellar parameters in these spectra, which in turn allows to analyze the star forming history of the Galaxy.

As more and more spectra are observed, and the census of stars in rarer classes grows, it may be possible to retrain the network to achieve an even better generalization across the entirety of the parameter space. However, the catalog presented here is a substantial step forward in characterizing spectroscopic stellar properties for all stars.

\software{ TOPCAT \citep{topcat}, APOGEE Net II}

\begin{acknowledgments}
KPR acknowledges support from ANID FONDECYT Iniciaci\'on 11201161. AS gratefully acknowledges funding support through Fondecyt Regular
(project code 1180350) and from the ANID BASAL project FB210003.  Funding for the Sloan Digital Sky Survey IV has been provided by the Alfred P. Sloan Foundation, the U.S. Department of Energy Office of Science, and the Participating Institutions. SDSS acknowledges support and resources from the Center for High-Performance Computing at the University of Utah. The SDSS web site is www.sdss.org.

SDSS is managed by the Astrophysical Research Consortium for the Participating Institutions of the SDSS Collaboration including the Brazilian Participation Group, the Carnegie Institution for Science, Carnegie Mellon University, Center for Astrophysics | Harvard \& Smithsonian (CfA), the Chilean Participation Group, the French Participation Group, Instituto de Astrofísica de Canarias, The Johns Hopkins University, Kavli Institute for the Physics and Mathematics of the Universe (IPMU) / University of Tokyo, the Korean Participation Group, Lawrence Berkeley National Laboratory, Leibniz Institut für Astrophysik Potsdam (AIP), Max-Planck-Institut für Astronomie (MPIA Heidelberg), Max-Planck-Institut für Astrophysik (MPA Garching), Max-Planck-Institut für Extraterrestrische Physik (MPE), National Astronomical Observatories of China, New Mexico State University, New York University, University of Notre Dame, Observatório Nacional / MCTI, The Ohio State University, Pennsylvania State University, Shanghai Astronomical Observatory, United Kingdom Participation Group, Universidad Nacional Autónoma de México, University of Arizona, University of Colorado Boulder, University of Oxford, University of Portsmouth, University of Utah, University of Virginia, University of Washington, University of Wisconsin, Vanderbilt University, and Yale University.
\end{acknowledgments}

\appendix
\section{Model Code} \label{sec:appendix}
\begin{verbatim}
class Model():

    class MetadataNet(nn.Module):
        """A simple feed-forward network for metadata."""

        def __init__(self):
            """Initializes Metadata Net as a 5-layer deep network."""

            super(Model.MetadataNet, self).__init__()
            self.l1 = nn.Linear(7, 8)
            self.l2 = nn.Linear(8, 16)
            self.l3 = nn.Linear(16, 32)
            self.l4 = nn.Linear(32, 32)
            self.l5 = nn.Linear(32, 64)

            self.activation = F.relu

        def forward(self, x):
            """Feeds some metadata through the network.
            
            Args:
                x: A minibatch of metadata.
            Returns:
                An encoding of the metadata to feed into APOGEE Net.
            """

            x = self.activation(self.l1(x))
            x = self.activation(self.l2(x))
            x = self.activation(self.l3(x))
            x = self.activation(self.l4(x))
            x = self.activation(self.l5(x))
            return x   


    class APOGEENet(nn.Module):

        def __init__(self, num_layers: int = 1, num_targets: int = 3, drop_p: float = 0.0) -> None:
            super(Model.APOGEENet, self).__init__()
            # 3 input channels, 6 output channels,  convolution
            self.conv1 = nn.Conv1d(num_layers, 8, 3, padding=1)
            self.conv2 = nn.Conv1d(8, 8, 3, padding=1)
            self.conv3 = nn.Conv1d(8, 16, 3, padding=1)
            self.conv4 = nn.Conv1d(16, 16, 3, padding=1)
            self.conv5 = nn.Conv1d(16, 16, 3, padding=1)
            self.conv6 = nn.Conv1d(16, 16, 3, padding=1)
            self.conv7 = nn.Conv1d(16, 32, 3, padding=1)
            self.conv8 = nn.Conv1d(32, 32, 3, padding=1)
            self.conv9 = nn.Conv1d(32, 32, 3, padding=1)
            self.conv10 = nn.Conv1d(32, 32, 3, padding=1)
            self.conv11 = nn.Conv1d(32, 64, 3, padding=1)
            self.conv12 = nn.Conv1d(64, 64, 3, padding=1)

            self.metadata = Model.MetadataNet()

            # an affine operation: y = Wx + b
            self.fc1 = nn.Linear(64*133*1 + 64, 512)
            self.fc1_dropout = nn.Dropout(p=drop_p)
            self.fc2 = nn.Linear(512, 512)
            self.fc3 = nn.Linear(512, num_targets)

        def forward(self, x: torch.Tensor, m: torch.Tensor) -> torch.Tensor:
            """Feeds data through the network.
            Args:
                x (Tensor): A spectra minibatch.
                m (Tensor): A metadata minibatch corresponding to x.
            Returns:
                A prediction from the network.
            """

            # Max pooling over a (2) window
            x = F.max_pool1d(F.relu(self.conv2(F.relu(self.conv1(x)))), 2)
            x = F.max_pool1d(F.relu(self.conv4(F.relu(self.conv3(x)))), 2)
            x = F.max_pool1d(F.relu(self.conv6(F.relu(self.conv5(x)))), 2)
            x = F.max_pool1d(F.relu(self.conv8(F.relu(self.conv7(x)))), 2)
            x = F.max_pool1d(F.relu(self.conv10(F.relu(self.conv9(x)))), 2)
            x = F.max_pool1d(F.relu(self.conv12(F.relu(self.conv11(x)))), 2)
            x = x.view(-1, self.num_flat_features(x))
            
            # Append outputs from the metadata
            m = self.metadata.forward(m)
            x = torch.hstack((x, m))
            x = F.relu(self.fc1_dropout(self.fc1(x)))
            x = F.relu(self.fc1_dropout(self.fc2(x)))
            x = self.fc3(x)
            return x

        def num_flat_features(self, x: torch.Tensor) -> int:
            """Returns the number of features in a flattened sample."""
            
            size = x.size()[1:]  # all dimensions except the batch dimension
            num_features = 1
            for s in size:
                num_features *= s
            return num_features

\end{verbatim}
\bibliography{main.bbl}

\end{document}